\documentclass[apjl]{emulateapj}
\usepackage{natbib}
\usepackage{color}
\lefthead{Jithesh and Wang}
\newcommand\chandra{{\it Chandra}}

\newcommand\xmm{{\it XMM-Newton}}
\newcommand\swift{{\it Swift}}

\newcommand\kms{\ifmmode {\rm~km\ s}^{-1} \else ~km s$^{-1}$\fi}
\newcommand\Hunit{\ifmmode {\rm~km\ s}^{-1}\ {\rm Mpc}^{-1}
        \else ~km s$^{-1}$ Mpc$^{-1}$\fi}
\newcommand\ctssec{\ifmmode {\rm~count\ s}^{-1} \else ~count s$^{-1}$\fi}
\newcommand\ergsec{\ifmmode {\rm~erg\ s}^{-1} \else
        ~erg s$^{-1}$\fi}
\newcommand\funit{\ifmmode {\rm~erg\ s}^{-1}\;{\rm cm}^{-2} \else
        ~ergs s$^{-1}$ cm$^{-2}$\fi}
\newcommand\phflux{\ifmmode {\rm~photon\ s}^{-1}\;{\rm cm}^{-2}
        \else   ~photon s$^{-1}$ cm$^{-2}$\fi}
\newcommand\efluxA{\ifmmode {\rm~erg\ s}^{-1}\;{\rm cm}^{-2}\;{\rm
        \AA}^{-1} \else ~erg s$^{-1}$ cm$^{-2}$ \AA$^{-1}$\fi}
\newcommand\efluxHz{\ifmmode {\rm~erg\ s}^{-1}\;{\rm cm}^{-2}\;{\rm
        Hz}^{-1} \else ~erg s$^{-1}$ cm$^{-2}$ Hz$^{-1}$\fi}
\newcommand\cc{\ifmmode {\rm~cm}^{-3} \else cm$^{-3}$\fi}
\newcommand\FWHM{\ifmmode {\rm~FWHM} \else ${\rm~FWHM}$\fi}
\newcommand\Msun{\ifmmode M_{\odot} \else $M_{\odot}$\fi}
\newcommand\Lsun{\ifmmode L_{\odot} \else $L_{\odot}$\fi}

\newcommand\hbeta{\ifmmode {\rm H}\beta \else H$\beta$\fi}
\newcommand\Kalpha{\ifmmode {\rm K}\alpha \else K$\alpha$\fi}
\newcommand\nh{\ifmmode N_{\rm H} \else N$_{\rm H}$\fi}

\usepackage{graphicx}

\begin{document}

\title{Transient X-ray source population in the Magellanic-type galaxy NGC 55}

\author{V. Jithesh and Zhongxiang Wang} 

\affil{Shanghai Astronomical Observatory, Chinese Academy of Sciences,\\ 
80 Nandan Road, Shanghai 200030, China; jithesh@shao.ac.cn}

\begin{abstract}

We present the spectral and temporal properties of 15 candidate transient 
X-ray sources detected in archival {\it XMM-Newton} and {\it Chandra} observations of 
the nearby Magellanic-type, SB(s)m galaxy NGC 55. On the basis of an X-ray 
color classification scheme, the majority of the sources may be identified 
as X-ray binaries (XRBs), and six sources are soft, including a likely 
supernova remnant. We perform a detailed spectral and variability analysis 
of the data for two bright candidate XRBs. Both sources displayed strong 
short-term X-ray variability, and their X-ray spectra and hardness ratios 
are consistent with those of XRBs. These results, combined with their high 
X-ray luminosities ($\sim 10^{38}$\,erg\,s$^{-1}$), strongly suggest that they are black 
hole (BH) binaries. Seven less luminous sources have spectral properties 
consistent with those of neutron star or BH XRBs in both normal and high-rate 
accretion modes, but one of them is the likely counterpart to a background 
galaxy (because of positional coincidence). From the spectral analysis, we 
find that the six soft sources are candidate super soft sources (SSSs), with 
dominant emission in the soft (0.3--2 keV) X-ray band. Archival {\it Hubble Space 
Telescope} optical images for 7 sources are available, and the data suggest 
that most of them are likely to be high-mass XRBs. Our analysis has revealed 
the heterogeneous nature of the transient population in NGC 55 (6 high-mass XRBs, 
1 low-mass XRBs, 6 SSSs, 1 active galactic nucleus), helping establish the 
similarity of the X-ray properties of this galaxy to those of other Magellanic-type galaxies.
\end{abstract}


\section{Introduction}
X-ray transients are an interesting class of sources, with quiescent 
luminosities below the detection limit, that are primarily discovered 
when they enter outbursts typically lasting from weeks to a few months. 
These bright outbursts are characterized by an episode of high accretion 
rates and abrupt increases of X-ray luminosity by several orders of magnitude. 
We now know that most of them are binary systems with a white dwarf (WD),
a neutron star (NS), or a black hole (BH) as the accreting compact object. 
Monitoring observations at X-ray wavelengths have revealed a variety of 
spectral and temporal behavior of these sources, and established distinct 
X-ray spectral states for X-ray binaries, in which the compact primary
is either a NS or a BH \citep[see][for extensive reviews on spectral 
states]{Mcc06, Rem06}. 

X-ray transient events in the nearby galaxies originate from different 
types of objects, which include ultraluminous X-ray sources (ULXs), 
super soft sources (SSSs), low-mass X-ray binaries (LMXBs) and 
high-mass X-ray binaries (HMXBs). Most of the ULXs identified in the nearby 
galaxies are persistent sources with X-ray luminosities ($> 10^{39}\ergsec$) 
exceeding the Eddington limit for a $\rm 10\ \Msun$ BH \citep[see][for a 
review]{Fen11}. Extensive monitoring observations with \xmm{}, \chandra{}, 
and \swift{} identified two ULX transients in M31 \citep{Kau12, Bar13}. 
Multi-wavelength analysis suggests that the underlying sources are likely 
stellar-mass BHs with low mass donors \citep{Noo12, Bar13}. Another 
important class of X-ray sources in the external galaxies are SSSs, 
usually characterized by values of $kT$ in the range of tens 
of eV, with X-ray luminosities between $10^{36}$ and 
$10^{38}\ergsec$ \citep{Dis03}. They were first identified in the Large 
Magellanic Cloud \citep[LMC;][]{Lon81}. A promising explanation for SSSs is 
the quasi-steady nuclear burning on the surface of a WD, in which the compact 
object accretes at a high rate from a Roche lobe-filling 
companion \citep{van92}. A sizable population of SSSs are identified in the 
external galaxies, for example M31, M101, M83, M51 and 
NGC 4697 \citep{Dis04, Sti11}, and some of these sources are sub-divided
into a new class called quasi-soft sources \citep[QSSs;][]{Dis04}. 
QSSs are sources with little or no emission above 2 keV, with temperatures 
significantly higher than that of the typical SSSs, generally between 100 eV 
and 350 eV. Detailed studies of SSSs revealed that they are detected in both 
early and late-type galaxies, identified in the field and globular clusters 
of a host galaxy, with a variety of temporal 
properties \citep{Dis03,Dis04a,Jen05,Bra12}. 
Finally, there will be possible contamination from background active galactic 
nuclei (AGN) in the X-ray source population of a galaxy, which are 
misidentified as the sources in the host galaxy \citep{Gut07, Gut13}. 

NGC 55 is a Magellanic-type, SB(s)m galaxy, a member of the nearby Sculptor 
group of galaxies. The first observation of NGC 55 at X-ray wavelengths 
by {\it ROSAT} \citep{Rea97, Sch97, Dah98} identified 15 discrete X-ray 
sources in the $\rm D_{25}$ ellipse of the galaxy. A detailed analysis of X-ray 
properties of NGC 55 was presented by \citet[][hereafter SRW06]{Sto06} 
using the \xmm{} observations. They identified 42 X-ray sources within 
the optical confines of the galaxy and classified them as X-ray 
binaries (XRBs), supernova remnants (SNRs), and SSSs. A ULX, with X-ray 
luminosity $> 10^{39}\ergsec$, is also present in this galaxy, which showed 
a significant variability with pronounced dips and a flux increase in the 
light curves during the \xmm{} observations \citep{Sto04}. However, from the {\it Swift} 
monitoring campaign and \chandra{} observations, \cite{Pin15} reported 
marginal evidence for limited dips in the light curves. 
Recently, \citet[][hereafter BWF15]{Bin15} presented a comprehensive 
study of X-ray point sources in three galaxies, which includes NGC 55, 
as a part of the Chandra Local Volume Survey (CLVS). 
The purpose of the CLVS was to identify the strong XRB candidates 
in nearby galaxies and compare the properties of the XRB population to the 
star formation histories of the host galaxies. They identified 154 X-ray 
sources in NGC 55 using multiple archival {\it Chandra} observations and 
studied the long-term variability using two {\it XMM-Newton} observations 
conducted in 2001. The X-ray hardness ratios, spectral 
properties, and temporal variability of the point sources were reported
and tentative classification of these sources was provided as well.
%%The main motivation of our study is to search the transient X-ray source population in NGC 55 and unveil the nature of these sources.  Here we analyze the long 127 ks {\it XMM-Newton} observation conducted in 2010 along with the other archival X-ray observations used in BWF15, to search transient X-ray source population in NGC 55.}  

%However, the observations used in our analysis overlap with BWF15, our main 
%motivation is to search the transient X-ray source population in NGC 55 and 
%unveil the nature of these sources.

As approximately one third of the known Galactic XRBs show evidence 
of being X-ray transients \citep{Tan96} but the extragalactic X-ray transient 
population has been relatively little studied (except in M31),
we have carried out studies of the transient X-ray source population in NGC 55.
Such studies for the external galaxies can give a better understanding 
of the nature and evolution of these sources. In this paper, we present 
our analysis of a long, 127 ks {\it XMM-Newton} observation conducted in 
2010 along with the other archival X-ray observations (also used in BWF15).
The spectral and temporal properties of transient X-ray sources identified 
in NGC 55 are reported. Using the {\it Hubble Space Telescope (HST)} 
observations and other multi-wavelength data, we address the possible 
nature of these sources. The paper is structured as follows. In \S 2, 
we describe the observations used and data reduction. We present the source 
selection and an overview of the source properties in \S 3. In \S 4, we 
discuss the results and possible nature of these sources, and in \S 5 
we summarize our results.

\section{Observations and Data Reduction}

\xmm{} observed NGC 55 three times: on 2001 November 14, 15, and 2010 May 25 
for exposure times of 33.6, 31.5, and 127.4 ks, respectively. The observations 
used in this work are summarized in Table \ref{obs}. The European Photon Imaging Camera (EPIC) 
PN \citep{Str01} and metal oxide semiconductor \citep[MOS;][]{Tur01} camera were
operated in the full frame mode using the thin optical blocking filter in the
2001 and 2010 observations. EPIC-PN and MOS data were reduced using standard 
tools ({\sc epchain} and {\sc emchain} respectively) of \xmm{} Science 
Analysis Software ({\sc sas}), version 14.0. The full-field background light 
curve was extracted to remove the particle flaring background. The good time 
intervals (GTI) were generated using periods with count rate $\le~0.8$ and 
$\le~0.3~\rm ct~s^{-1}$ in the 10--12 keV light curve for PN and MOS data 
respectively. We selected the events corresponding to patterns 0--4 from 
the PN data and patterns 0--12 from the MOS data for the analysis 
with (FLAG == 0). 

\begin{table}
\tabletypesize{\small}
\tablecolumns{4}
\setlength{\tabcolsep}{5.0pt}
\tablewidth{320pt}
	\caption{Observation log}
 	\begin{tabular}{@{}lccc@{}}
	\hline
\colhead{Data} & \colhead{Obs ID} & \colhead{Date} & \colhead{Exposure Time$^a$} \\
\hline

%\begin{deluxetable}{lccc}
%\tabletypesize{\small}
%\setlength{\tabcolsep}{5.0pt}
%\tablewidth{320pt}
%\tablecaption{Observation log}
%\startdata
%\hline
%\hline
%Data & Obs ID & Date & Exposure Time$^a$ \\
%\hline

XMM1 & 0028740201 & 2001 Nov 14 & 33.6 \\
XMM2 & 0028740101 & 2001 Nov 15 & 31.5 \\
XMM3 & 0655050101 & 2010 May 25 & 127.4 \\
{\it Chandra} & 2255     & 2001 Sep 11 & 60.1 \\
{\it Chandra} & 4744     & 2004 Jun 29 & 9.7 \\
{\it HST} DISK & 9765    & 2003 Dec 16 & 0.7 (F814W) \\
 	       &         &             & 0.7 (F606W) \\
{\it HST} FIELD & 9765   & 2003 Sep 23 & 0.7 (F814W) \\
		 &       &             & 0.4 (F606W) \\
{\it HST} WIDE-1 & 11307 & 2007 Jul 28 & 2.6 (F814W) \\
		 &       &             & 0.9 (F606W) \\
{\it HST} WIDE-3 & 11307 & 2007 Aug 07 & 3.9 (F814W) \\
		 &       &             & 2.7 (F606W) \\
{\it HST} WIDE-4 & 11307 & 2007 Aug 09 & 3.9 (F814W) \\
		 &       &             & 2.7 (F606W) \\
{\it HST} WIDE-5 & 11307 & 2007 Aug 06 & 3.9 (F814W) \\
		 &       &             & 2.7 (F606W) \\
%\enddata 
\hline
\end{tabular} 
\tablecomments {$^a$ Exposure time is in units of kiloseconds.}
\label{obs}
%\end{deluxetable}
\end{table}

The source detection routine was carried out in the high sensitive EPIC-PN 
data over the entire energy band. The SAS task {\sc eboxdetect} was used 
to perform the initial source detection 
(with a detection threshold `likemin'= 8), employing a sliding box detection 
method. We provided the obtained source positions as input for 
the task {\sc esplinemap} that constructs background maps using source-free 
regions of the image. The {\sc eboxdetect} task was performed again in `map' 
mode using the available background map, which improved detection 
sensitivity. We ran the {\sc emldetect} task using these improved background 
maps with a minimum detection likelihood value of 10 (mlmin=10). If this 
value is 10 or greater, the sources were classified as significant detection 
and included in the final source list. Using the same criteria, we obtained 
the final source list from the 2001 and 2010 observations. The X-ray sources 
in the final list were then astrometrically corrected by correlating with 
the USNO A2.0 optical catalog \citep{Mon98}, in which 
the SAS task {\sc eposcorr} was used.

We also analyzed two archival {\it Chandra} observations of NGC 55, conducted 
in 2001 September 11 and 2004 June 29 with the Advanced CCD Imaging 
Spectrometer Imaging Array (ACIS-I). However, the second observation was 
too short and none of any transient sources were detected in it. Thus we 
report no further analysis of this observation in the paper. 
The {\it Chandra} data were reduced and reprocessed using the science threads 
of {\it Chandra} Interactive Analysis of Observations software 
package ({\sc ciao}) version 4.6 and HEASOFT version 6.15.1. 

In addition, we checked the archival \swift{} X-Ray Telescope 
\citep[XRT;][]{Bur05} observations for possible detection of the sources
reported in this paper. However the sources, 
except T1 \citep[See][hereafter JW15]{Jit15}, were too faint 
to be detected with the \swift{} XRT telescope. 

The Optical analysis was carried out using the six {\it HST} fields (See 
Table \ref{obs} for details). We searched the candidate optical counterparts 
of transient sources after performing an astrometric calibration of the 
X-ray and {\it HST} images using the bright point sources from the Two Micron 
All Sky Survey (2MASS) Source Catalog \citep{Skr06}. We computed the plate 
solution using the {\tt IRAF} task {\tt ccmap} and the root-mean-square (rms) 
residuals obtained from {\tt ccmap} are typically less than few hundredths of an 
arcsecond in both R.A and Decl. Moreover, the total alignment error, computed by 
summing the X-ray and optical rms residuals in quadrature, are much smaller than 
the X-ray positional uncertainties. Thus, we continued the search for optical 
counterparts to the transient sources using the positional uncertainties quoted 
in the Table \ref{tc}.

\section{Analysis and Results}

\subsection{Transient Source Selection}

Transient X-ray sources were identified by visual inspection as well as 
comparing the final source lists from the 2001 and 2010 \xmm{} observations. 
We selected the sources which `disappeared' or `appeared' in the long, 
127 ks \xmm{} observation and classified them as transient candidate (TC) or 
possible transient candidate (PTC). These sources were detected in at least 
one observation with an unabsorbed 0.3--8 keV luminosity of 
$> 1\times 10^{36} \ergsec$ at a $> 4\sigma$ confidence level and not detected 
in another observation. In addition, we calculated the luminosity ratios 
between the ``on-state" (the peak X-ray luminosities) and ``off-state" 
(the non-detection upper limits) for these sources to confirm their
transient nature. We classified the sources with the ratios $> 5$ as TCs and 
those with the ratios between 1 and 5 as PTCs. The flux ratio used here is 
slightly different from the value used (high/low ratios $\sim 8$) in 
the \chandra{} studies of transient sources in M33 \citep{Wil08}. 
The classification leads to the detection of 15 TCs and PTCs in 
NGC 55, listed in Table~\ref{tc}. 

\begin{table*}
\centering
\tabletypesize{\small}
\tablecolumns{10}
\setlength{\tabcolsep}{9.0pt}
\tablewidth{320pt}
	\caption{Properties of the transient candidate sources detected in NGC 55}
 	\begin{tabular}{@{}lccccccccr@{}}
	\hline
\colhead{Src} & \colhead{Catalog} & \colhead{ObsID} & \colhead{R.A.} & \colhead{Dec.} & \colhead{$r_{1\sigma}$} & \colhead{Extr. Rad} & \colhead{Net} & \colhead{Ratio} & \colhead{TC/PTC} \\
\colhead{No.} & \colhead{No.} & & (h:m:s) & (${^\circ}:':''$) & (arcsec) & (arcsec) & \colhead{Counts} & & \\
\hline

%\begin{deluxetable}{lccccccccr}
%\centering
%\tabletypesize{\footnotesize}
%\setlength{\tabcolsep}{4.0pt}
%\tablecaption{Properties of the transient candidate sources detected in NGC 55}
%\startdata
%\hline
%\hline
%Src & Catalog & ObsID & R.A. & Dec. & $r_{1\sigma}$ & Extr. Rad & Net & Ratio & TC/PTC \\
%No. & No. & & (h:m:s) & (${^\circ}:':''$) & (arcsec) & (arcsec) & Counts & & \\
%\hline

T1 & T1(JW15) & XMM3 & 00:14:46.81 & --39:11:23.48 & 0.54 & 14 & 2134 & 10.7 & TC \\
T2 & 75(SRW06) & XMM1 & 00:15:34.27 & --39:14:24.80 & 0.52 & 15 & 3009 & 95.5 & TC \\
T3 & 27(SRW06) & XMM2 & 00:14:35.45 & --39:11:32.20 & 1.50 & 13 & 66 & 1.3 & PTC \\
T4 & 30(SRW06) & XMM2 & 00:14:37.18 & --39:11:22.00 & 2.26 & 14 & 69 & 1.1 & PTC \\
T5 & 37(SRW06) & XMM2 & 00:14:43.74 & --39:12:41.50 & 1.49 & 12 & 73 & 1.3 & PTC \\
T6 & 44(SRW06) & XMM1 & 00:14:52.68 & --39:12:23.90 & 1.26 & 20 & 103 & 4.1 & PTC \\
T7 & -- & XMM3 & 00:14:54.86 & --39:14:16.97 & 1.79 & 20 & 87 & 1.0 & PTC \\
T8 & 51(SRW06) & XMM1 & 00:15:00.62 & --39:12:16.40 & 1.16 & 14 & 95 & 1.8 & PTC \\
T9 & 53(SRW06) & XMM1 & 00:15:03.79 & --39:14:59.80 & 1.52 & 15 & 45 & 6.5 & TC \\
T10 & -- & XMM3 & 00:15:13.74 & --39:13:27.12 & 1.65 & 15 & 154 & 1.6 & PTC \\
T11 & -- & XMM3 & 00:15:21.64 & --39:16:12.46 & 2.16 & 13 & 105 & 1.7 & PTC \\
T12 & -- & XMM3 & 00:15:43.89 & --39:17:55.22 & 0.96 & 15 & 136 & 5.9 & TC \\
T13 & -- & XMM3 & 00:15:52.85 & --39:16:34.39 & 1.81 & 20 & 94 & 5.0 & TC \\
T14 & -- & XMM3 & 00:16:11.32 & --39:16:37.05 & 1.20 & 15 & 188 & 5.8 & TC \\
T15 & 122(SRW06) & XMM1 & 00:16:19.34 & --39:16:45.50 & 2.61 & 15 & 57 & 3.9 & PTC \\

%\enddata
\hline
\end{tabular}
\tablecomments {(1) Source number used in this paper; (2) Source identification number in JW15 and SRW06; 
(3) Observation ID in which the source was detected; (4)-(5) Right Ascension (R.A.) and Declination (Dec.) of each source (J2000.0); (6) $1\sigma$ positional uncertainty of the source; (7) Extraction radius; (8) Net counts from the EPIC-PN and MOS in 0.3--8 keV; (9) Luminosity Ratio; (10) Source Classification as TC or PTC (see text for details).}
\label{tc}
%\end{deluxetable}
\end{table*}

SRW06 identified 42 X-ray sources in the $\rm D_{25}$ region of NGC~55 using 
the XMM1 and XMM2 observations. We identified nine more sources in the XMM3 
observation. Out of nine sources, two (XMMU J001548.1-391612 and XMMU J001604.6-391538) 
of them were not cataloged in SRW06, but detected in the XMM1 and XMM2 
observations. Thus these two sources do not satisfy our criteria and are 
not included in the analysis. The remaining seven sources in XMM3 were again 
verified by performing the source detection procedure in the 0.3-1 keV, 1-2 keV, 
and 2-6 keV images. These sources were detected with a $4\sigma$ threshold 
(mlmin=10) in at least one of the three energy bands. Moreover, 
they were cataloged in the third generation XMM-Newton Serendipitous Source 
Catalog (3XMM-DR4)\footnote{http://xmmssc-www.star.le.ac.uk/Catalogue/3XMM-DR4/} 
and flagged as the sources detected with high quality (summary flag $\leq 1$, 
where a low summary flag value indicates a high quality for a 
detection\footnote{http://xmmssc-www.star.le.ac.uk/Catalogue/2XMM/UserGuide\_xmmcat.html}). 
In addition, there are eight sources identified in SRW06 but not detected
in the XMM3 observation (See Table~\ref{tc}). In total, our classification 
identified six TCs and nine PTCs in NGC 55. 
Among them, T1, whose properties have been reported in JW15 in detail,
is likely a transient black hole XRB in the star-forming region of NGC~55.
For completeness, the source is also listed in this paper.

\subsection{Hardness Ratio and Variability}
\label{subsec:hr}

Since transient sources have few net counts, hardness ratios (HRs) can be 
considered as a primary tool to investigate their spectral properties. 
Although we cannot conclusively classify an individual source based on 
its X-ray colors alone, however it is possible to identify the trend in 
the source population. Therefore, the HRs of the transients were calculated 
from the count rates, which are defined as HR1=(M$-$S)/(M+S) and HR2=(H$-$M)/(H+M), 
where S, M and H are the count rates in soft (0.3--1 keV), medium (1--2 keV), 
and hard (2--6 keV) bands respectively. For the sources detected in EPIC-PN 
and MOS camera, we quote the weighted mean of hardness ratios (weights 
were calculated using the count rates) and they are listed in Table~\ref{hr}. 
The X-ray color classification scheme of \cite{Jen05} developed for 
{\it XMM-Newton} was used to classify the transient sources. The scheme 
divides the X-ray sources into four broad categories: absorbed sources (ABSs; HR1 $> 0.57$), 
XRBs ($-0.24 <$ HR1 $< 0.57$, $-0.8 <$ HR2 $< 0.8$), SNRs (HR1 $< -0.24$, HR2 $< -0.10$) 
and background sources (HR1 $< -0.24$, HR2 $> -0.10$; See Table 3 of \cite{Jen05} 
for more details). If we combine the `absorbed' and `XRB' 
categories into a single `XRB' category, as in SRW06, nine out of 15 sources 
fall in the XRB category. The remaining six sources are soft, and one 
of them, T10, falls under the SNR category.
 
Five transient sources were detected in the 2001 \chandra{} observation 
and we classified them using the color classification scheme of 
\cite{Kil05}, tuned for \chandra{} data (See Table 2 of \cite{Kil05} for more details). 
The obtained HRs for these sources are consistent with those from the \xmm{} observations. 

%XMM == 5-XRB, 4-Absorbed, 5-soft, 1-snr
%chandra == 3-XRB, 2-Absorbed

\begin{figure}
\centering
\includegraphics[width=8.5cm,height=8.5cm,angle=0]{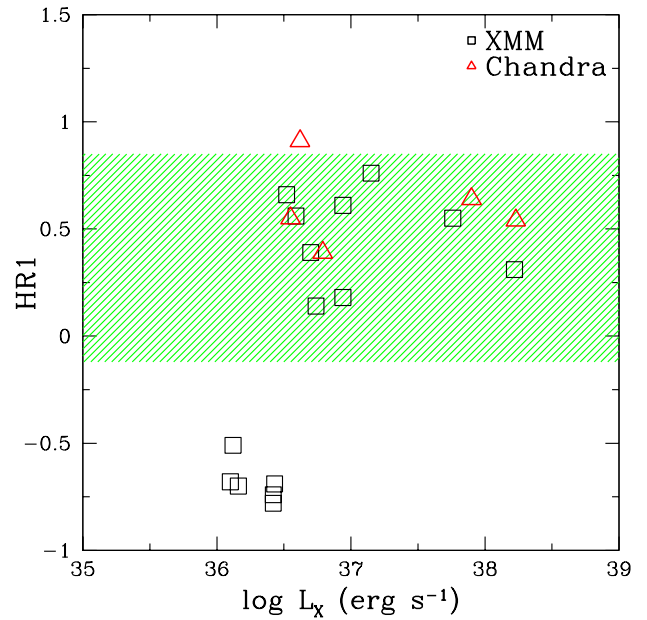}
\caption{Hardness ratio versus X-ray luminosity for the transient candidates in NGC 55. The uncertainties are not plotted in order to have a clear view of the data points, which are given in Table~\ref{hr}. The shaded region indicates the range of persistent X-ray binary sources in the $\rm D_{25}$ region of the galaxy. The X-ray luminosities were estimated from best-fit single component models (see Section \ref{subsec:spec})}
\label{lxhr}
\end{figure}

The HR--luminosity diagram (hardness--intensity proxy) of the TCs/PTCs 
detected in NGC 55 is given in Figure~\ref{lxhr}. From the figure it is clear 
that the majority of the TCs/PTCs have the color consistent with those of 
the persistent XRBs in the $\rm D_{25}$ region of the galaxy. For six source, 
their HR values are softer than the persistent XRB population and 
X-ray luminosities are lower compared to the other transient candidates, 
which suggest that they are the soft source population in this galaxy. 

\begin{table}
\tabletypesize{\small}
\tablecolumns{6}
\setlength{\tabcolsep}{4.0pt}
\tablewidth{240pt}
	\caption{Hardness ratios and luminosities of transient candidates}
 	\begin{tabular}{@{}lccccr@{}}
	\hline
\colhead{Src} & \colhead{ObsID} & \colhead{HR1} & \colhead{HR2} & \colhead{Class} & \colhead{$\rm L_{X}$}  \\
\colhead{No.} & & & & & \\
\hline

%\begin{deluxetable}{lccccr}
%\tabletypesize{\small}
%\setlength{\tabcolsep}{6.0pt}
%\tablewidth{320pt}
%\tablecaption{Hardness ratios and luminosities of transient candidates}
%\startdata
%\hline
%\hline
%Src  & ObsID  & HR1  & HR2  & Class & $\rm L_{X}$ \\
%No. & & & & & \\
%\hline

T1 & XMM3 & $0.55\pm0.03$ & $-0.37\pm0.02$ & XRB & $37.76^{+0.03}_{-0.03}$ \\%dbb
T2 & XMM1 & $0.31\pm0.01$ & $-0.09\pm0.01$ & XRB & $38.22^{+0.02}_{-0.02}$ \\%pl
   & C-2255 & $0.54\pm0.01$ & $-0.06\pm0.01$ & XRB & $38.23^{+0.03}_{-0.03}$ \\%pl
T3 & XMM2 & $-0.78\pm0.30$ & $0.27\pm0.93$ & SOFT & $36.42^{+0.40}_{-0.30}$ \\ %pl
T4 & XMM2 & $-0.70\pm0.26$ & $-0.17\pm0.29$ & SOFT & $36.16^{+0.20}_{-0.35}$ \\ %pl
T5 & XMM2 & $0.66\pm0.35$ & $-0.31\pm0.09$ & ABS & $36.52^{+0.21}_{-0.22}$ \\ %pl
   & C-2255 & $0.55\pm0.11$ & $-0.31\pm0.07$ & XRB & $36.55^{+0.18}_{-0.18}$ \\ %pl
T6 & XMM1 & $0.76\pm0.50$ & $-0.16\pm0.05$ & ABS & $37.15^{+0.34}_{-0.19}$ \\ %pl
   & C-2255 & $0.91\pm0.14$ & $-0.11\pm0.02$ & ABS & $36.62^{+0.15}_{-0.14}$ \\ %pl
T7 & XMM3 & $-0.68\pm1.70$ & $0.08\pm1.59$ & SOFT & $36.10^{+0.14}_{-0.19}$ \\ %dbb
T8 & XMM1 & $0.18\pm0.07$ & $-0.13\pm0.17$ & XRB & $36.94^{+0.23}_{-0.20}$ \\ %pl
   & C-2255 & $0.39\pm0.09$ & $0.30\pm0.05$ & XRB & $36.79^{+0.13}_{-0.13}$ \\ %pl
T9 & XMM1 & $-0.69\pm0.30$ & $-0.33\pm0.31$ & SOFT & $36.43^{+0.20}_{-0.20}$ \\%pl
T10 & XMM3 & $-0.51\pm0.22$ & $-0.73\pm1.26$ & SNR & $36.12^{+0.26}_{-0.34}$ \\ %pl
T11 & XMM3 & $0.61\pm0.78$ & $0.29\pm0.12$ & ABS & $36.94^{+2.71}_{-0.58}$ \\ %pl
   & C-2255 & $0.64\pm0.15$ & $0.33\pm0.07$ & ABS & $37.90^{+1.52}_{-1.23}$ \\%pl
T12 & XMM3 & $0.14\pm0.02$ & $-0.16\pm0.03$ & XRB & $36.74^{+0.38}_{-0.17}$ \\ %pl 
T13 & XMM3 & $0.56\pm0.22$ & $0.06\pm0.01$ & ABS & $36.59^{+0.91}_{-0.25}$ \\ %pl
T14 & XMM3 & $-0.74\pm0.10$ & $-0.76\pm0.43$ & SOFT & $36.42^{+0.12}_{-0.12}$ \\%pl 
T15 & XMM1 & $0.39\pm0.25$ & $-0.30\pm0.15$ & XRB & $36.70^{+2.10}_{-0.38}$ \\ %pl

%\enddata
\hline
\end{tabular} 
\tablecomments {(1) Source number; (2) Observation ID, where C indicates the {\it Chandra} observation used in the analysis; (3)--(4) Hardness Ratios derived from the count rate (See \S 3.2); (5) Nature of the source according to the classification scheme of \cite{Jen05} and \cite{Kil05}; (6) Unabsorbed 0.3--8 keV X-ray luminosity in units of $\ergsec$ derived from best-fit single component model.}
\label{hr}
%\end{deluxetable}
\end{table}

For the transient sources T1 and T2, we extracted the background subtracted light curve based on the combined EPIC-PN and MOS camera over 0.3--8 keV energy range. We investigated the short-term X-ray variability by performing the Kolmogorov--Smirnov (K-S) test on the light curves binned with 800-s, 300-s and 100-s. From the K-S test, we found that these sources showed a strong short-term X-ray variability at confidence level of $> 99.99\%$. 

%(See Figure 2 of JW15 for T1 and Figure \ref{lc} of this paper for T2) 

%The X-ray luminosity and HR of persistent sources are taken from SRW06.

\subsection{Spectral Analysis}
\label{subsec:spec}

X-ray spectra of the transient sources within the $\rm D_{25}$ region of NGC 55 
were extracted in the 0.3--8 keV band, using a circular extraction aperture 
with radius ranges from 12--20 arcsec. The different extraction 
radii were used to avoid the contamination from the nearby sources in 
the crowded region. Background spectra were extracted 
from nearby source-free regions, with the same aperture radius. 
We used the {\sc sas} task {\sc especget} to obtain source and background spectra 
together with the ancillary response file (ARF) and redistribution matrix 
file (RMF) required in the spectral analysis. We binned the data to a minimum 
of 20 counts per bin and in many cases $\chi^2$ minimization was used to fit the data. 
In other cases where there were no sufficient spectral counts, the Cash 
Statistic \citep{Cas79} was used for spectral fitting. Because the Cash Statistic 
does not provide a goodness-of-fit measure like $\chi^2$, we additionally 
performed 5000 Monte Carlo simulations of each spectrum using the {\tt Goodness} task \citep {Arn96} 
to evaluate the quality of the fit for Cash Statistic. The fit is considered to be 
acceptable, if the ``goodness'' is $\le 50\%$. For $\chi^2$ statistic, we used 
the null hypothesis probability to check the quality. 
The spectra were fitted in the {\sc xspec} version 12.8.1g \citep{Arn96}. 
The unabsorbed fluxes were derived using the convolution model {\tt CFLUX} of {\sc xspec} 
and the X-ray luminosities were estimated by assuming the distance of 
1.78 Mpc \citep{Kar03}. All errors quoted were computed at a 90\% confidence 
level.

We initially fitted both EPIC-PN and MOS spectra simultaneously using 
single-component models such as power law ({\tt PL}) and multicolor disc 
blackbody ({\tt DISKBB}). In all cases, the preferred model is the model 
with the lowest $\chi^2/\rm d.o.f$. 
However, favoring the spectral models for 
the low counting statistics sources is difficult, hence we report only 
the absorbed power law model parameters for these sources. 
An intervening absorption column 
({\tt TBABS}) was also applied to each model, which includes the Galactic 
column density towards NGC 55. If the best-fit value of $\rm N_{H}$ is below 
the Galactic absorption, we froze $\rm N_{H}$ to the Galactic value, 
$1.72\times10^{20}~\rm cm^{-2}$ \citep{Dic90}. The spectral fitting results are summarized 
in Table~\ref{fit}. For the majority (13 out of 15 sources) of the sources a power law provides a marginally or 
significantly better fit than a disk blackbody model.

\begin{table*}
\tabletypesize{\small}
\tabletypesize{\footnotesize}
\setlength{\tabcolsep}{4.5pt}
\tablewidth{42pc}
	\caption{Best-fit parameters of the source spectra}
 	\begin{tabular}{@{}lcccccccccccr@{}}
	\hline
\colhead{Src} & & & & \colhead{$\rm PL^a$} & & & & \colhead{$\rm DISKBB^a$} & & \\	
\colhead{No}  & \colhead{ObsID$^b$} & \colhead{$\rm N_{H}$$^c$} & \colhead{$\Gamma^d$} & \colhead{$\rm log~L_{X}$$^e$} & \colhead{$\rm \chi^2/\rm d.o.f^f$} & \colhead{$\rm G^g$} & \colhead{$\rm N_{H}$$^c$} & \colhead{$\rm kT_{in}$$^h$} & \colhead{$\rm log~L_{X}$$^e$} & \colhead{$\rm \chi^2/\rm d.o.f^f$} & \colhead{$\rm G^g$} & \colhead{$\rm log~L_{XUP}$$^i$} \\
\hline

%\begin{deluxetable}{lcccccccccccr}
%\tabletypesize{\footnotesize}
%\setlength{\tabcolsep}{1.5pt}
%\tablewidth{43pc}
%\tablecaption{Best-fit parameters of the source spectra}
%\startdata
%\hline
%\hline
%Src & & & & $\rm PL^a$ & & & & $\rm DISKBB^a$ & & & \\	
%No  & ObsID$^b$ & $\rm N_{H}$$^c$ & $\Gamma^d$ & $\rm log~L_{X}$$^e$ & $\rm \chi^2/\rm d.o.f^f$ & $\rm G^g$ & $\rm N_{H}$$^c$ & $\rm kT_{in}$$^h$ & $\rm log~L_{X}$$^e$ & $\rm \chi^2/\rm d.o.f^f$ & $\rm G^g$ & $\rm log~L_{XUP}$$^i$ \\    
%\hline

T1 & XMM3 & $0.68^{+0.07}_{-0.06}$ & $3.15^{+0.18}_{-0.17}$ & $38.38^{+0.11}_{-0.10}$ & $110.7/105$ & 0.33 & $0.31^{+0.04}_{-0.04}$ & $0.72^{+0.05}_{-0.05}$ & $37.76^{+0.03}_{-0.03}$ & $97.4/105$ & 0.69 & $< 36.73$ \\
T2 & XMM1 & $0.19^{+0.03}_{-0.03}$ & $1.71^{+0.09}_{-0.09}$ & $38.22^{+0.02}_{-0.02}$ & $183.5/150$ & 0.03 & $0.04^{+0.02}_{-0.02}$ & $1.58^{+0.12}_{-0.11}$ & $38.10^{+0.02}_{-0.02}$ & $210.2/150$ & 0.001 & $< 36.24$ \\
 & C-2255 & $0.19^{+0.06}_{-0.06}$ & $1.56^{+0.13}_{-0.13}$ & $38.23^{+0.03}_{-0.03}$ & $71.4/70$ & 0.43 & $0.03^{+0.04}_{-0.03}$ & $1.79^{+0.23}_{-0.19}$ & $38.12^{+0.02}_{-0.03}$ & $77.1/70$ & 0.26 & \\
T3 & XMM2 & $0.0172(f)$            & $4.29^{+2.68}_{-1.49}$ & $36.42^{+0.40}_{-0.30}$ & $6.4/6\rm (C)$ & 27\% & & & & & & $< 36.32$ \\
T4 & XMM2 & $0.0172(f)$            & $3.40^{+1.68}_{-0.98}$ & $36.16^{+0.20}_{-0.35}$ & $9.3/7\rm (C)$ & 44\% & & & & & & $< 36.12$ \\
T5 & XMM2 & $0.0172(f)$            & $1.69^{+0.67}_{-0.55}$ & $36.52^{+0.21}_{-0.22}$ & $7.3/6\rm (C)$ & 37\% & & & & & & $< 36.41$ \\
 & C-2255 & $0.0172(f)$              & $1.51^{+0.82}_{-0.72}$ & $36.55^{+0.18}_{-0.18}$ & $4.4/4\rm (C)$ & 22\% & & & & & &  \\
T6 & XMM1 & $0.24^{+0.42}_{-0.24}$ & $1.71^{+1.23}_{-0.81}$ & $37.15^{+0.34}_{-0.19}$ & $14.4/13\rm (C)$ & 32\% & & & & & & $< 36.54$ \\
 & C-2255 & $0.0172(f)              $ & $1.69^{+0.96}_{-0.77}$ & $36.62^{+0.15}_{-0.14}$ & $5.6/5\rm (C)$ & 21\% & & & & & &  \\
T7 & XMM3 & $0.0172(f)$            & $4.82^{+1.77}_{-1.23}$ & $36.17^{+0.14}_{-0.19}$ & $32.5/25$ & 0.15 & $0.0172(f)$            & $0.10^{+0.05}_{-0.03}$ & $36.10^{+0.14}_{-0.19}$ & $31.5/25$ & 0.17 & $< 36.09$ \\ 
T8 & XMM1 & $0.10^{+0.31}_{-0.10}$ & $1.61^{+1.00}_{-0.59}$ & $36.94^{+0.23}_{-0.20}$ & $10.7/11\rm (C)$ & 16\% & & & & & & $< 36.69$ \\
 & C-2255 & $0.0172(f)              $ & $0.90^{+0.45}_{-0.46}$ & $36.79^{+0.13}_{-0.13}$ & $4.3/8\rm (C)$ & 6\% & & & & & & \\
T9 & XMM1 & $0.0172(f)$            & $5.64^{+3.70}_{-1.76}$ & $36.43^{+0.20}_{-0.20}$ & $4.5/11\rm (C)$ & 1\% & & & & & & $< 35.62$ \\
T10 & XMM3 & $0.0172(f)$            & $2.08^{+1.11}_{-0.75}$ & $36.12^{+0.26}_{-0.34}$ & $17.0/21$ & 0.71 & $0.0172(f)$           & $0.45^{+0.46}_{-0.23}$ & $35.93^{+0.20}_{-0.28}$ & $17.1/21$ & 0.71 & $< 35.92$ \\ 
T11 & XMM3 & $2.50^{+4.20}_{-1.94}$ & $2.36^{+3.41}_{-1.56}$ & $36.94^{+2.71}_{-0.58}$ & $6.1/9$ & 0.73 & $1.62^{+2.94}_{-1.28}$ & $1.61^{+0.00}_{-0.98}$ & $36.54^{+0.39}_{-0.30}$ & $6.6/9$ & 0.68 & $< 36.72$ \\ 
 & C-2255 & $3.82^{+5.05}_{-2.23}$ & $3.42^{+1.34}_{-2.32}$ & $37.90^{+1.52}_{-1.23}$ & $2.5/3\rm (C)$ & 11\% & & & & & & \\
T12 & XMM3 & $0.32^{+0.31}_{-0.21}$ & $2.08^{+1.13}_{-0.71}$ & $36.74^{+0.38}_{-0.17}$ & $8.7/11$ & 0.65 & $0.12^{+0.19}_{-0.12}$ & $1.21^{+0.98}_{-0.53}$ & $36.54^{+0.14}_{-0.16}$ & $9.5/11$ & 0.58 & $< 35.97$ \\
T13 & XMM3 & $0.48^{+0.99}_{-0.44}$ & $1.51^{+2.38}_{-1.17}$ & $36.59^{+0.91}_{-0.25}$ & $26.9/14$ & 0.02 & $0.27^{+0.67}_{-0.25}$ & $2.86^{+0.00}_{-2.30}$ & $36.53^{+0.20}_{-0.31}$ & $27.6/14$ & 0.02 & $< 35.89$ \\
T14 & XMM3 & $0.0172(f)$            & $3.78^{+1.17}_{-0.91}$ & $36.42^{+0.12}_{-0.12}$ & $12.0/10$ & 0.28 & $0.0172(f)$            & $0.12^{+0.09}_{-0.04}$ & $36.31^{+0.14}_{-0.16}$ & $16.7/10$ & 0.08 & $< 35.66$ \\ 
T15 & XMM1 & $0.27^{+1.35}_{-0.27}$ & $2.08^{+5.73}_{-1.21}$ & $36.70^{+2.10}_{-0.38}$ & $0.6/3\rm (C)$ & 1\% & & & & & & $< 36.11$ \\

%\enddata
\hline
\end{tabular}
\tablecomments {$^{a}$Spectral models used for fitting: PL---power law continuum; DISKBB---multi-color disc blackbody. $^{b}$Observations used in each fit. $^{c}$Absorption column density, including Galactic absorption, in units of $10^{22}~\rm cm^{-2}$. $^d$Power law index. $^e$Logarithmic unabsorbed 0.3--8 keV X-ray luminosity in units of $\rm erg~s^{-1}$, calculated by assuming the distance of 1.78 Mpc \citep{Kar03}. $^f$The $\chi^2/\rm d.o.f$ value for the model, If C-statistics is adopted, it is indicated with a C. $^g$Goodness of fit (Goodness when using Cash-statistics and null hypothesis probability for $\chi^2$ statistics). $^h$Inner disc temperature in units of keV. $^i$ Upper limit on the X-ray luminosity for non-detection.}
\label{fit}
%\end{deluxetable}
\end{table*}

We estimated the upper limits on the count rates using {\sc eregionanalyse} 
task in {\sc sas} for the sources that were not detected in 
the {\it XMM-Newton} observations. After accounting for the background, 
the 90\% confidence upper limits in 0.3--8 keV energy band were derived. 
Assuming for each source the best-fit spectral model derived from 
the observations with detection, the flux upper limits were also obtained. 

\begin{figure}
\centering
\includegraphics[width=6.5cm,height=7.5cm,angle=-90]{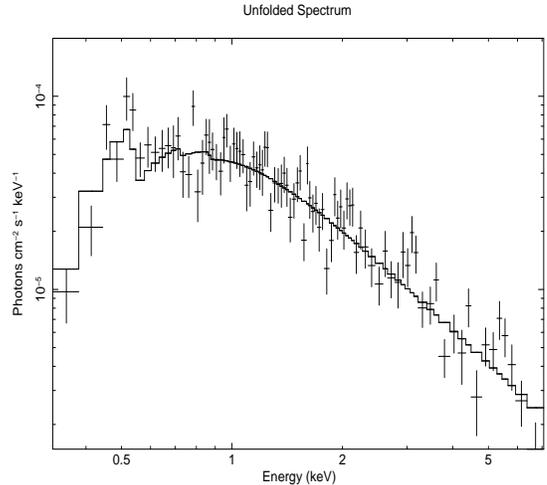}
\caption{ \xmm{} EPIC-PN 0.3--8 keV spectrum of T2, fitted with an absorbed 
power law model.}
\label{spectrum}
\end{figure}

Source T2 has 3009 and 1781 net counts detected in the \xmm{} and \chandra{} 
observations respectively. A single-component model, absorbed power law, 
provided an acceptable fit (See Figure \ref{spectrum}) with 
$\Gamma=1.71\pm0.09$, $\chi^2/\rm d.o.f = 183.5/150$ for the \xmm{} and 
$\Gamma=1.56\pm0.13$, $\chi^2/\rm d.o.f = 77.1/70$ for the \chandra{} observations.
The implied 0.3--8 keV luminosity from our best-fit model 
is $1.70^{+0.12}_{-0.11} \times 10^{38}~\rm erg~s^{-1}$. We also attempted a 
two-component model fitting, power law plus disc blackbody, for T2. 
The two-component model provided a marginal improvement to the spectral fit, 
with $\Delta \chi^2 \sim 4$ for two extra degrees of freedom, over the 
single-component fit. The best fit yielded power-law index $\Gamma = 1.79\pm0.15$ plus 
$0.14^{+0.09}_{-0.03}~\rm keV$ disc blackbody, absorbed by 
$(0.36\pm0.16) \times 10^{22}~\rm cm^{-2}$, with $\chi^2/\rm d.o.f = 179.8/148$ 
and the unabsorbed luminosity is $L_{X} = 2.74^{+3.15}_{-1.01} \times 10^{38}~\rm erg~s^{-1}$. 
The marginal improvement over the single component fit indicates that
the presence of the second component is tentative, and we do not have a good constraint 
on the flux contribution from the disc. The estimated disc blackbody component contribution
to the total 0.3--8 keV source flux is $< 35\%$.

\cite{Bar14} successfully classified the X-ray transient sources in the galaxy
M31 using the double thermal model (disc blackbody and blackbody), which 
was proposed by \cite{Lin07} to examine the spectral evolution of transient 
LMXBs. \cite{Bar14} identified 36 BH candidates that exhibit apparent hard state 
spectra with luminosities much higher than that for NS LMXBs. Using 
the double thermal model parameters, they found that none of the BH candidates 
occupied in the NS LMXB region. We also tested the double thermal model 
for T2 and the spectral fit is not significantly improved 
($\chi^2/\rm d.o.f = 182.9/149$, $\Delta \chi^2 \sim 1$ for one extra degree of 
freedom) over the absorbed power law model, but the spectral parameters, 
$kT_{in} = 0.54\pm0.12$ keV, $kT_{BB} = 1.30^{+0.29}_{-0.15}$ keV, 
and $L_{X} ({\rm 2-10~keV}) = 1.2^{+0.66}_{-0.55} \times 10^{37}~\rm erg~s^{-1}$ 
are consistent with the case of BH candidates. 

%\begin{figure}
%\centering
%\includegraphics[width=8.5cm,height=8.5cm,angle=0]{f2.eps}
%\caption{Combined EPIC-PN and MOS 0.3--8 keV light curve of T2. The light curve has been background subtracted and binned with 300-s.}
%\label{lc}
%\end{figure}

%(\#8, \#10, \#44, \#51, \#122, \#4 and \#6)
%(T12, T13,    T6,  T8,   T15,  T10 and T11)

%\#37 -> T5

Seven sources (T5, T6, T8, T11, T12, T13, and T15) have spectra marginally 
or well fitted with the absorbed power law with $\Gamma \sim 1.6 - 2.4$ and the 
model luminosities are in the range of $ \sim 3.3 - 14.1 \times 10^{36}~\rm erg~s^{-1}$.
For the disc blackbody model, the inner disc temperature obtained 
ranges from $\sim 1.2 - 2.9$ keV for the sources T11, T12 and T13.  
In most of the cases, the values of $\rm N_{H}$ obtained higher than 
the Galactic foreground column density towards NGC 55. These spectral analysis suggests that these
source are consistent with being NS and BH binaries in either a hard state or 
a thermal dominated state \citep{Rem06}. 

%\#4 -> T10

In the section \ref{subsec:hr}, we found that source T10 had HRs consistent 
with those of SNRs. We fitted its spectrum with a thermal plasma model, 
{\tt APEC} in {\sc xspec}. The spectrum is best-fitted with a thermal plasma 
temperature of $0.17^{+0.24}_{-0.13}$ keV, a solar metal abundance and 
an absorption of $0.77^{+1.15}_{-0.51} \times10^{22}~\rm cm^{-2}$ 
($\chi^2/\rm d.o.f = 18.5/20$). The estimated unabsorbed 0.5--2 keV X-ray 
luminosity is $\sim 3.3 \times 10^{37}~\rm erg~s^{-1}$. We also tried 
the blackbody ({\tt BBODY}) model for this source and the obtained
parameters are given in Table~\ref{bbody}.

%soft sources
%(\#53, \#13, \#27, \#30 and \#5)
%(T9,    T14,  T3,   T4  and  T7)

\begin{figure}
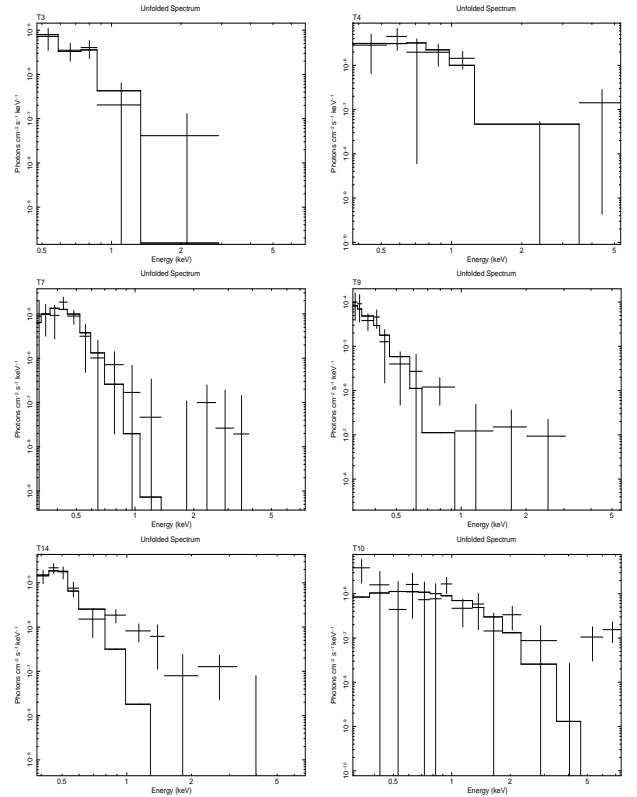

\centering
\includegraphics[width=3.5cm,height=4.1cm,angle=-90]{f3a.eps}
\includegraphics[width=3.5cm,height=4.1cm,angle=-90]{f3b.eps}
\includegraphics[width=3.5cm,height=4.1cm,angle=-90]{f3c.eps}
\includegraphics[width=3.5cm,height=4.1cm,angle=-90]{f3d.eps}
\includegraphics[width=3.5cm,height=4.1cm,angle=-90]{f3e.eps}
\includegraphics[width=3.5cm,height=4.1cm,angle=-90]{f3f.eps}
\caption{\xmm{} EPIC-PN 0.3--8 keV spectra of candidate soft sources (T3, T4, T7, T9, T14 and T10), 
fitted with an absorbed blackbody model. The blackbody fit becomes worse compared to 
the power law for T14. Most of them have only a little or no emission above 
2 keV.}
\label{sss}
\end{figure}

The sources that are classified as soft sources by the HRs 
(T3, T4, T7, T9 and T14) have spectra favored by the power law as well as 
disc blackbody model. In all cases, the power law model fits yielded steep power 
law indices ($\Gamma \sim 3.4 - 5.6$), reflecting the soft nature of 
the emission. Moreover, the inner disc temperature of these sources obtained 
from the disc blackbody model fits are much softer 
($kT_{in} \sim 0.1 - 0.12$ keV) than that of the typically observed 
Galactic XRBs ($kT_{in} \sim 0.7 - 2$ keV). For these source, we tested 
the {\tt BBODY} model and the spectral parameters are given in 
Table~\ref{bbody}. The spectral fits obtained from the {\tt BBODY} model are 
very similar to the power law or disc blackbody, except for T14. For 
the source T14, the {\tt BBODY} fit is worse compared to the power law 
(See Figure \ref{sss}). The temperature yielded from the {\tt BBODY} model 
for these sources are in the range of $\sim 50 - 180$ eV, consistent with that
of SSSs, and the X-ray luminosities are $\sim 10^{36}~\rm erg~s^{-1}$.

We note that among the 154 X-ray sources in NGC 55 identified by BWF15, 
five sources (T2, T5, T6, T8 and T11) of ours were also listed.
The rest of the transient sources were not included in BWF15 
and half of them were detected in the 2010 {\it XMM-Newton} observation. 
We compared the results of these five sources with BWF15 and their hardness 
ratio classification are consistent with our results. 
They have studied the spectral properties of X-ray sources with $> 50$ net 
counts from \chandra{} data and only one source (T2) is listed in our transient 
catalog. Their best-fit model for this source is an absorbed power law and
the best-fit parameters are $N_{H} = 0.15\pm0.06\times 10^{22}\rm cm^{-2}$ and
$\Gamma = 1.5\pm0.1$ (the unabsorbed $\rm L_{X}$ in 0.3--8 keV band 
is $\sim 2\times 10^{38} \rm erg~s^{-1}$), consistent with that of ours. 
They have identified long-term variations for four sources listed in this work 
(T2, T5, T6 and T11) from \chandra{} and \xmm{} observations, which are 
confirmed by our analysis presented here. 

\begin{table}
\tabletypesize{\small}
\tablecolumns{6}
\setlength{\tabcolsep}{6.0pt}
\tablewidth{240pt}
	\caption{Spectral parameters for super soft sources using {\tt BBODY} model}
 	\begin{tabular}{@{}lccccr@{}}
	\hline
\colhead{Src No.} & \colhead{$\rm N_{H}$} & \colhead{$\rm kT$} & \colhead{$\rm log~L_{X}$} & \colhead{$\rm \chi^2/\rm d.o.f$} & \colhead{$\rm G$} \\
\hline

%\begin{deluxetable}{lccccr}
%\tabletypesize{\small}
%\setlength{\tabcolsep}{6.0pt}
%\tablewidth{280pt}
%\tablecaption{Spectral parameters for super soft sources using {\tt BBODY} model}
%\startdata
%\hline
%\hline
%Src No. & $\rm N_{H}$ & $\rm kT$ & $\rm log~L_{X}$ & $\rm \chi^2/\rm d.o.f$ & $\rm G$ \\
%\hline

T3  & $1.72(f)$           & $120^{+51}_{-38}$ & $36.17^{+0.22}_{-0.26}$ & $4.9/6$ & 13\% \\ %dbb
T4 & $1.72(f)$           & $176^{+52}_{-45}$ & $36.08^{+0.17}_{-0.25}$ & $4.4/7$ & 6\% \\ %dbb
T7 & $1.72(f)$           & $78^{+30}_{-23}$ & $36.08^{+0.15}_{-0.19}$ & $31.4/25$ & 0.18 \\ %dbb
T9 & $1.72(f)$           & $50^{+44}_{-20}$ & $36.41^{+0.24}_{-0.26}$ & $6.2/11$ & 2\% \\%pl
T14 & $1.72(f)$           & $90^{+37}_{-22}$ & $36.31^{+0.14}_{-0.16}$ & $17.5/10$ & 0.06 \\%pl 

T10 & $1.72(f)$           & $229^{+190}_{-85}$ & $35.83^{+0.19}_{-0.27}$ & $18.0/21$ & 0.65 \\ %dbb

%\enddata 
\hline
\end{tabular} 
\tablecomments {(1) Source number; (2) Absorption column density in $10^{20}~\rm cm^{-2}$; (3) Black body temperature in eV; (4) Logarithmic unabsorbed 0.3--8 keV X-ray luminosity in $\rm erg~s^{-1}$; (5) The $\chi^2/\rm d.o.f$ values for the model; (6) Goodness of fit.}
\label{bbody}
%\end{deluxetable}
\end{table}

\subsection{Optical Analysis and Comparison with Multi-wavelength Catalogs}

The transient sources were cross-correlated with the multi-wavelength 
catalogs available in the NED\footnote{http://ned.ipac.caltech.edu/}, 
VizieR\footnote{http://vizier.u-strasbg.fr/viz-bin/VizieR/}, 
and SIMBAD\footnote{http://simbad.u-strasbg.fr/simbad/} data bases. 
We searched these catalogs for matches within the $3\sigma$ error radius of 
each \xmm{} position. Of the 15 transient sources, source T15 positionally 
coincides with a galaxy, LCRS B001349.4-393325 \citep{She96} and is 
likely a background AGN (SRW06). BWF15 reported the counterpart for 
T2 and T11 from {\it GALEX} GR6 data release. However, 
these {\it GALEX} sources are not in the $3\sigma$ error circles of the 
\xmm{} positions we have derived and hence we do not consider them associated 
to any of the X-ray sources. 

%{\bf This difference is mainly due to the strict criteria 
%used to search the counterpart (i.e., $3\sigma$ error circles of the 
%\xmm{} positions).}

\begin{figure*}
%\begin{center}
%\centering
%  \subfloat[NGC 4486]{\label{fig:NGC 4486}\includegraphics[width=0.28\textwidth]{f1a.eps}}     

\includegraphics[width=4.2cm,height=4.2cm]{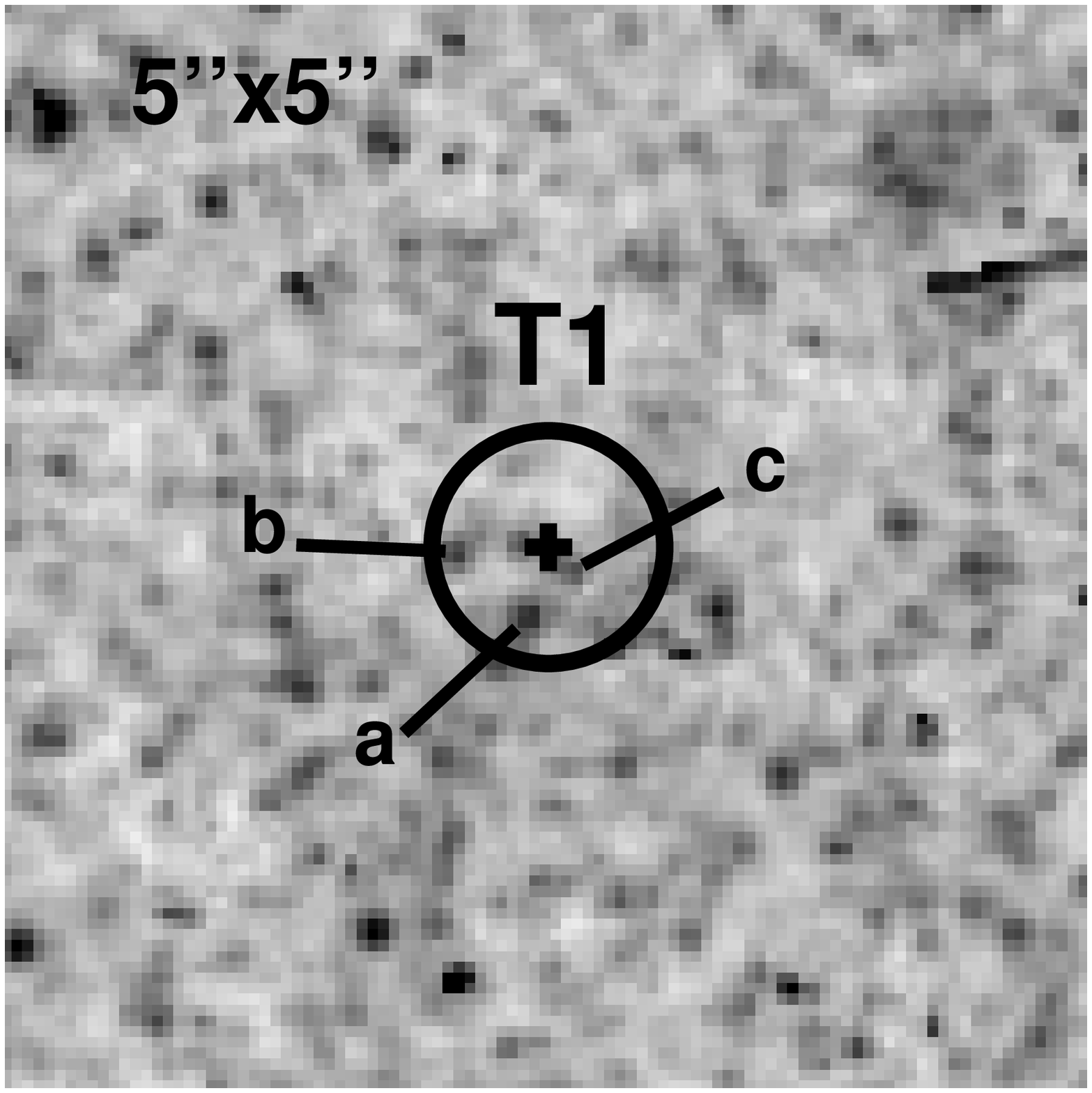}
\includegraphics[width=4.2cm,height=4.2cm]{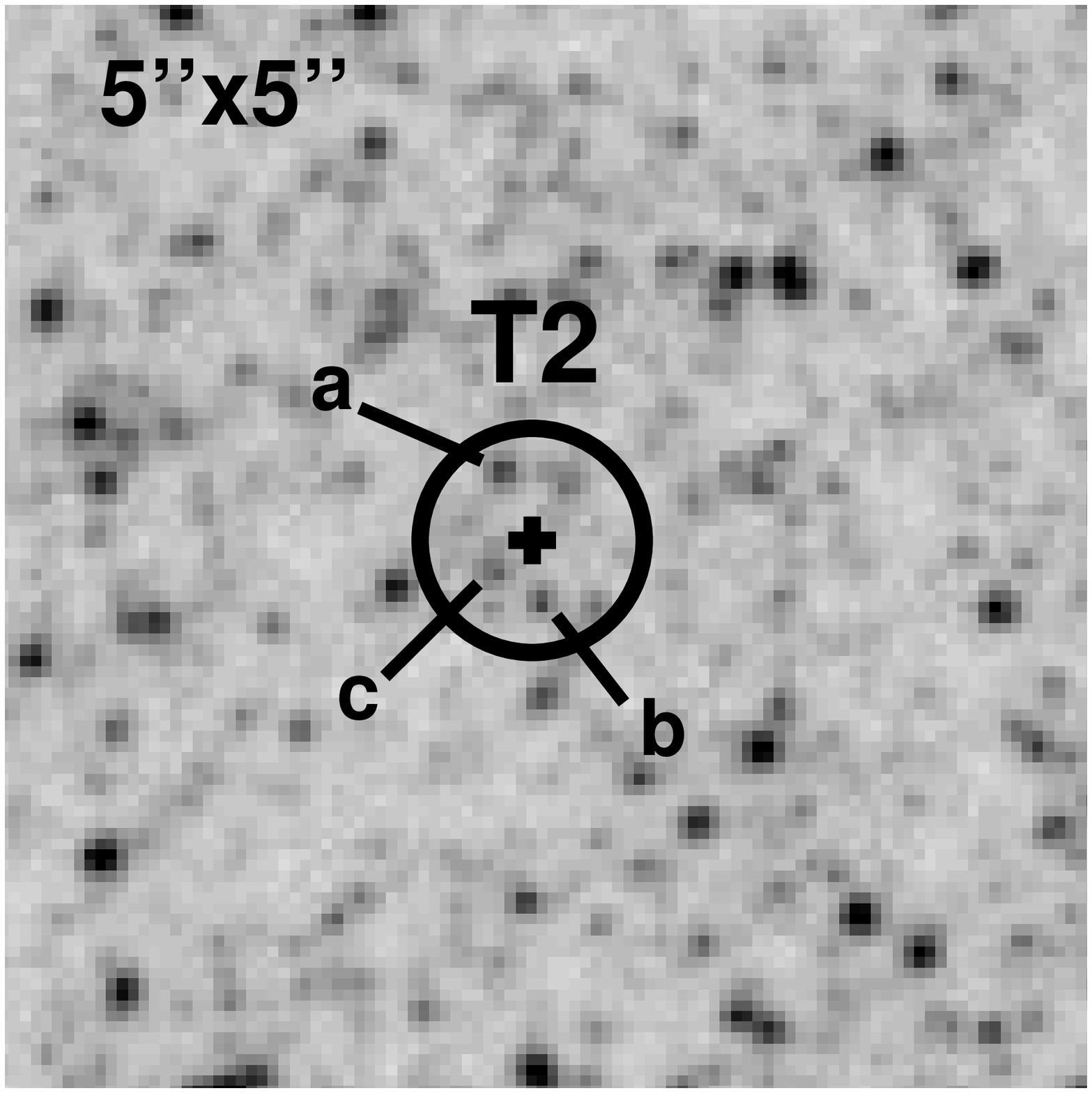}
\includegraphics[width=4.2cm,height=4.2cm]{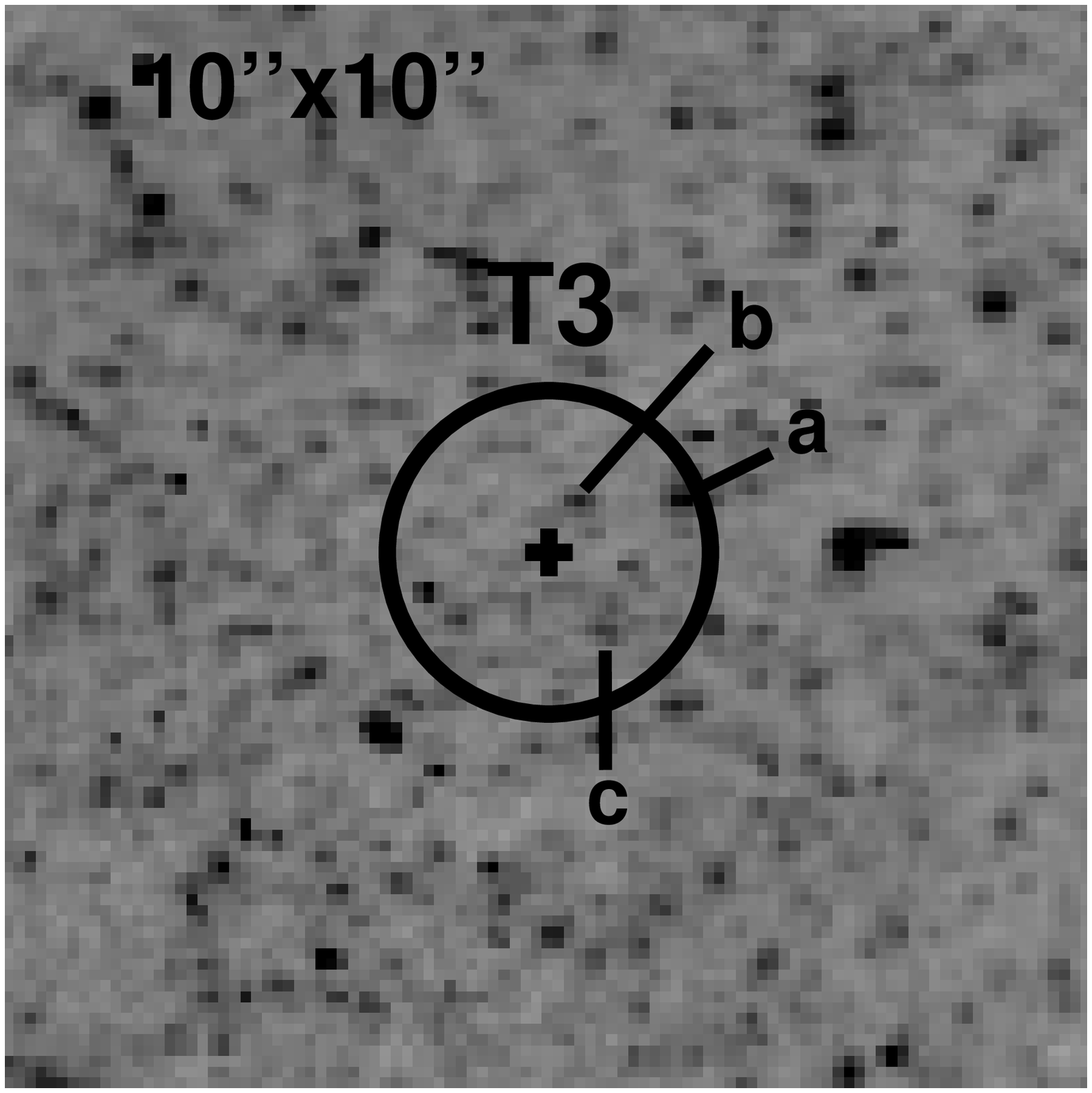}
\includegraphics[width=4.2cm,height=4.2cm]{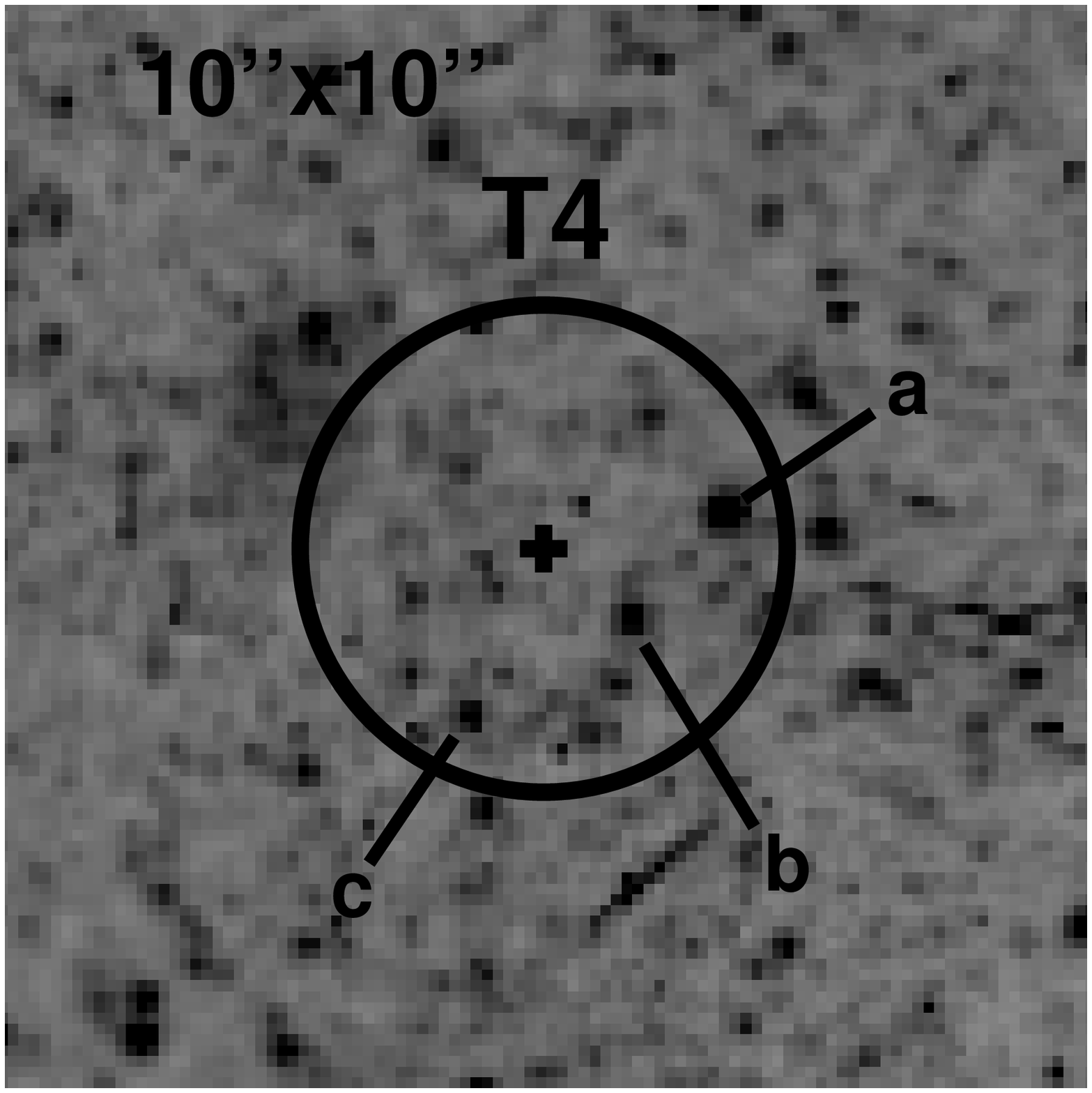}

\includegraphics[width=4.2cm,height=4.2cm]{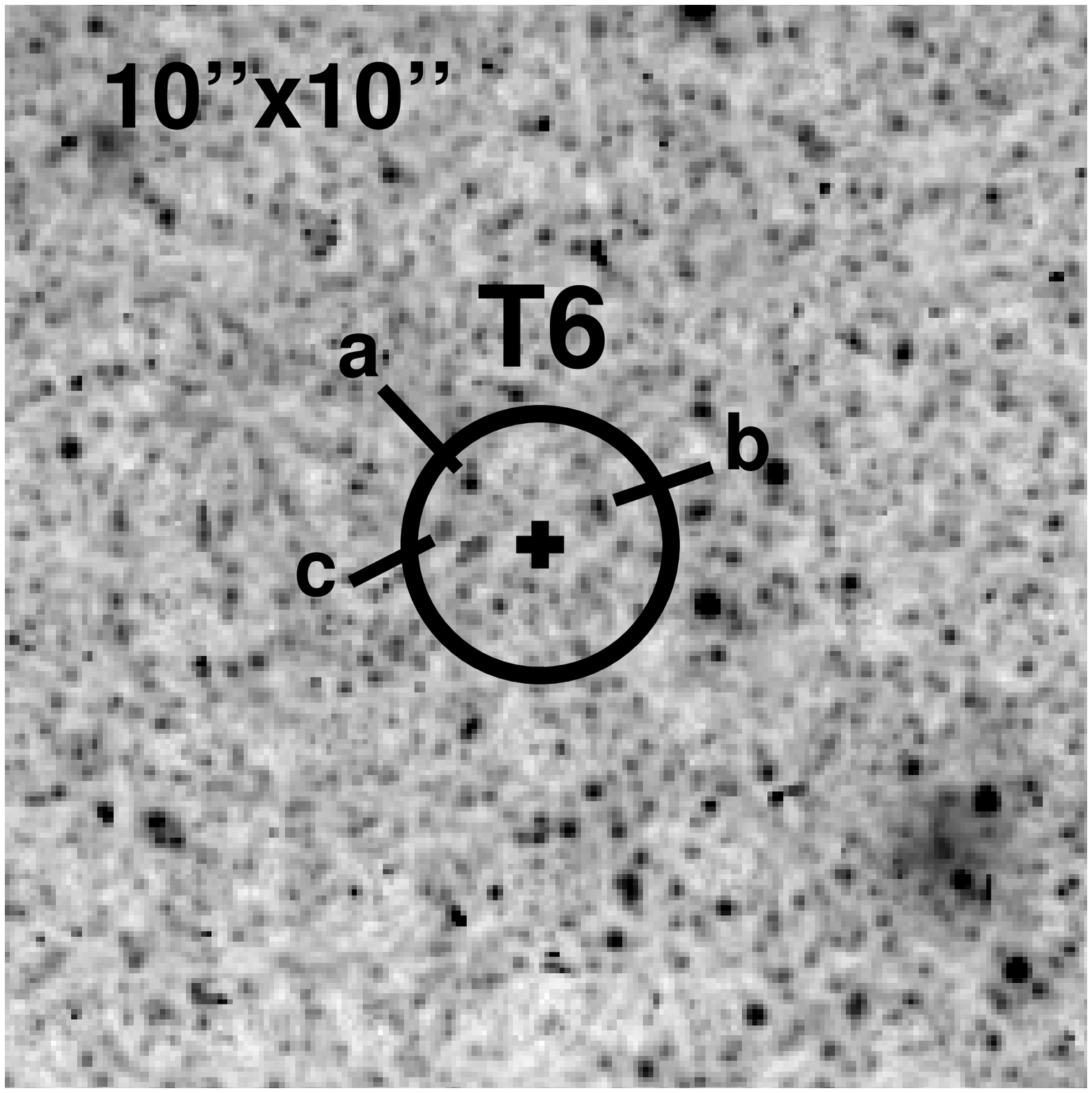}
\includegraphics[width=4.2cm,height=4.2cm]{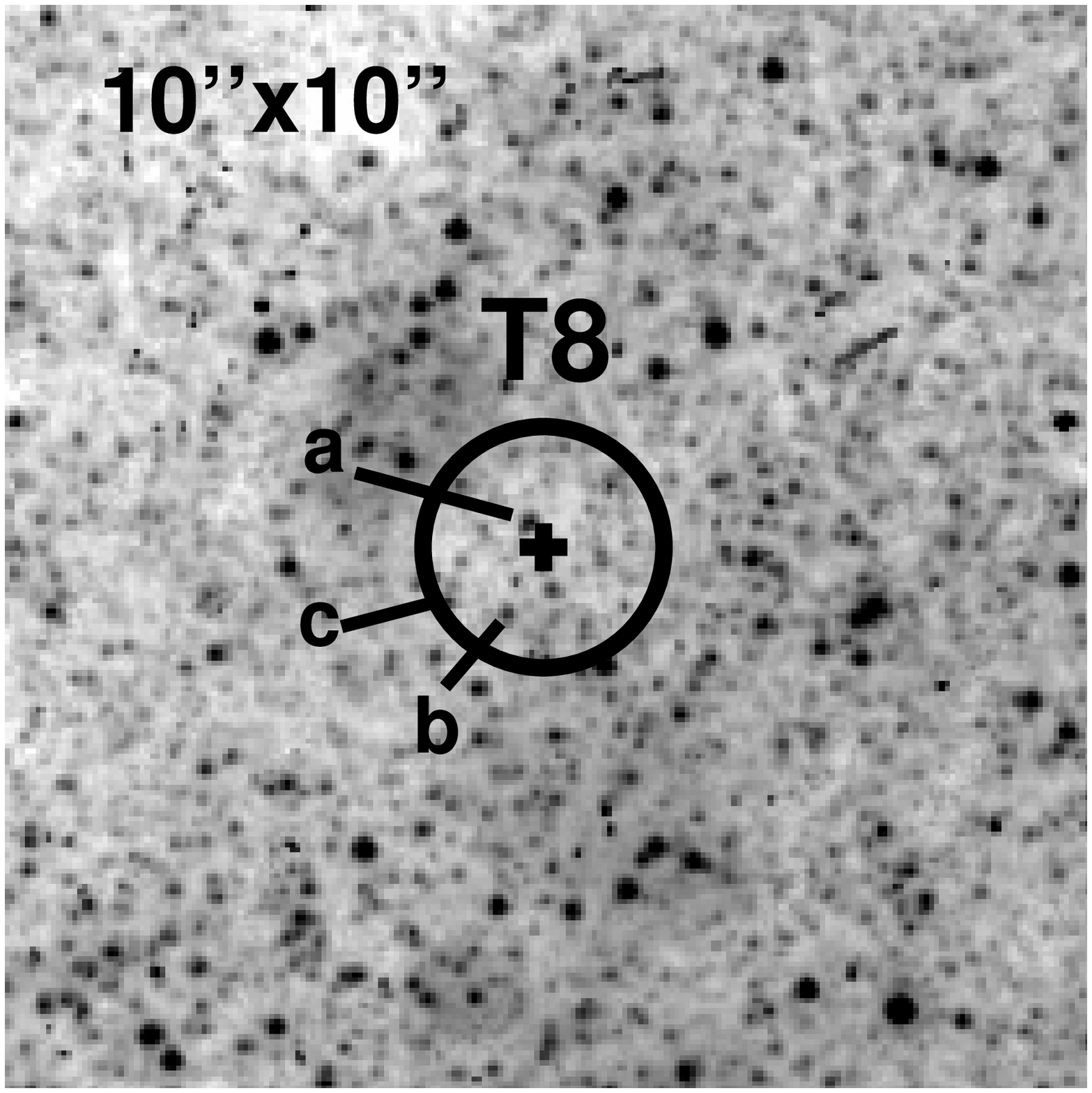}
\includegraphics[width=4.2cm,height=4.2cm]{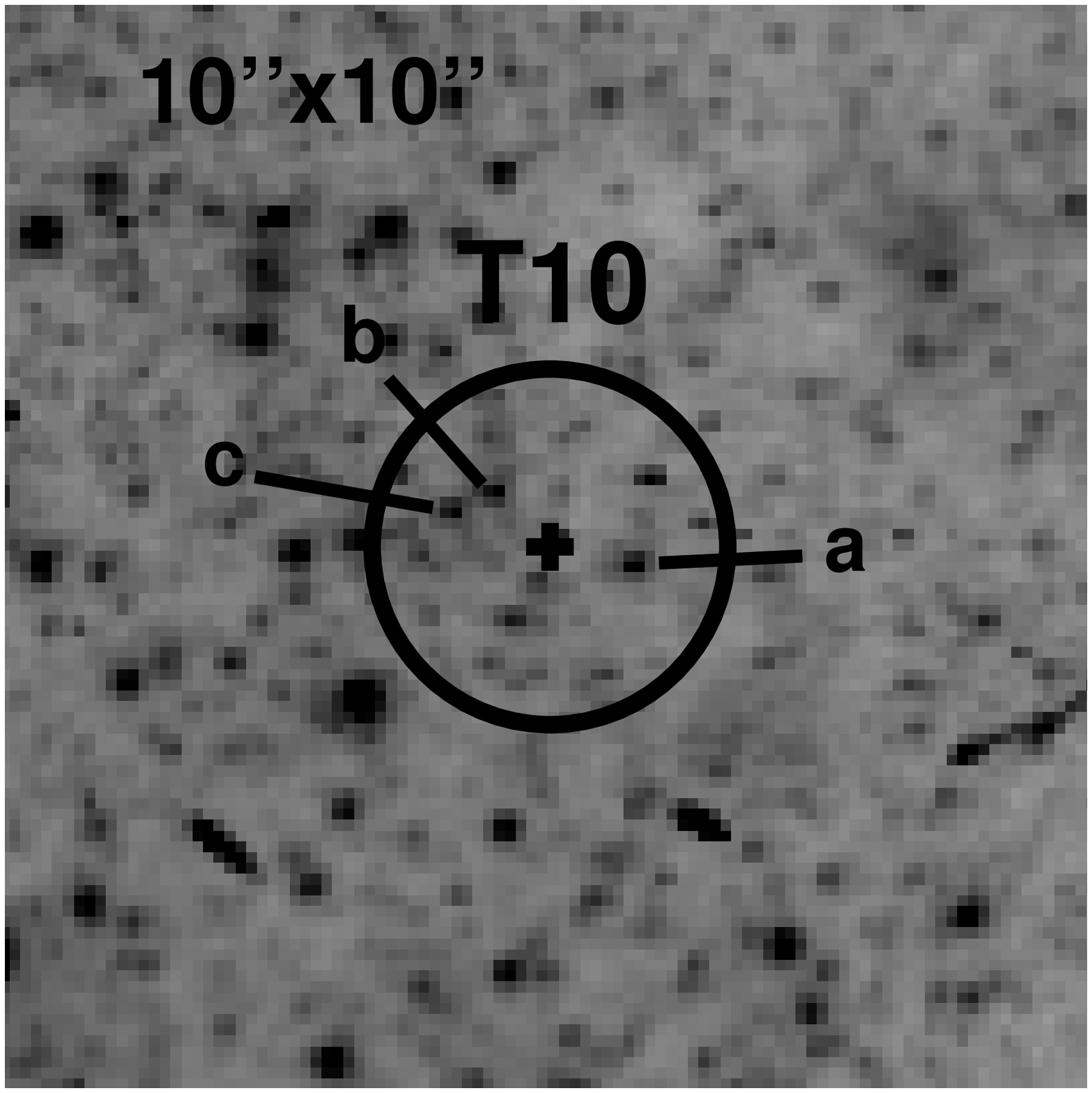}

\caption{{\it HST V}-band images for the transient X-ray sources in NGC 55. The box size ($5''\times5''$ or $10''\times10''$) is given in the left-top corner of each image. The black circle shows the $1\sigma$ positional error circle centered on the source position, indicated by the 'plus' sign. The three brightest objects inside the error circle were labeled as "a", "b", "c" and their absolute $V$ magnitude and color are given in Table \ref{class}. North is up, and east is to the left.}
\label{hst}
%\end{center}
\end{figure*}

We searched possible counterparts for the transient sources in 
the {\it HST} observations. Out of the 15 TCs/PTCs sources, seven are covered 
in the fields of view of {\it HST}. For all these sources, the source region 
is crowded and the \xmm{} error circle itself contains multiple optical 
sources. The {\it HST} fields are shown in Figure \ref{hst}. We computed 
the magnitudes of sufficiently bright sources in each error circle 
using {\tt DAOPHOT} \citep{Ste87} package in {\tt IRAF} in F814W ($I$) and F606W ($V$) bands.
The magnitudes and colors, not corrected for reddening, of
the brightest objects are given in Table~\ref{class}. The absolute $V$ 
magnitudes and $V-I$ colors of the optical sources have the values of 
$\sim$ [$-$0.2, $-$5.1] and $\sim$ [$-$0.8, 1.5] respectively, which are 
broadly consistent with blue/red bright giants or supergiants, or with background AGNs. 
We also estimated the logarithmic X-ray--to--optical flux ratio $\log(f_{x}/f_{o}$), 
where $f_{x}$ is the 0.3--8 keV flux and $f_{o}$ is the F606W flux. 

In addition to the HRs and X-ray spectral fitting, we added
X-ray--to--optical flux ratios and multi-wavelength information to 
characterize the possible nature of the transient sources. Table \ref{class} 
summarizes the X-ray, optical and multi-wavelength observations available 
for each of the transient sources. Stars exhibit a wide range of flux ratio 
values, but typically $\log(f_{x}/f_{o}) < 0$. For AGNs and BL Lac objects, 
their flux ratios are $> 1$ \citep{Mac82,Sto91}. Moreover, the known accreting 
X-ray pulsars and HMXBs in the Small Magellanic Cloud (SMC) have a flux ratio 
$\lesssim 1$ with $B-V \lesssim 0$ \citep{Mcg08}. If one of the optical sources 
in an errorcircle is the counterpart, the flux ratios would suggest that T1 and T2 
are an AGN, and rest of them are possible HMXBs. However, T1 exhibited outburst 
activity where accretion might occur only during certain parts of an orbit, 
which could fit in with a high mass donor star. Moreover, the colors of the 
optical sources in its error circle are not very blue in nature. If one 
of them is the counterpart of T1, it is more consistent with those of 
types A or F bright giant stars. Source T2 also has possible blue optical 
counterpart. BWF15 classified this source as an LMXB from their analysis using
color-magnitude diagrams, X-ray color-magnitude diagrams,
and the X-ray--to--optical flux ratio. Therefore the optical data 
suggest most of them are HMXBs, and for T1 and T2, their overal 
properties favor XRB nature.

\begin{table*}
\centering
\tabletypesize{\small}
\tablecolumns{9}
\setlength{\tabcolsep}{10.0pt}
\tablewidth{420pt}
	\caption{X-ray, optical and multi-wavelength data for the transient candidate sources detected in NGC 55}
 	\begin{tabular}{@{}lcccccccr@{}}
	\hline
\colhead{Src} & \colhead{HR} & \colhead{Variability} & \colhead{X-ray} & \colhead{$\rm M_{V}$} &  \colhead{$V-I$} & \colhead{Range of} & \colhead{$\log(f_{x}/f_{o}$)} & \colhead{Multi-} \\
\colhead{No.} & \colhead{Class} & & Spectrum & & & $V-I$ & & wavelength \\
\hline

%\begin{deluxetable}{lcccccccr}
%\tabletypesize{\small}
%\setlength{\tabcolsep}{4.0pt}
%\tablewidth{420pt}
%\tablecaption{X-ray, optical and multi-wavelength data for the transient candidate sources detected in NGC 55}
%\startdata
%\hline
%\hline
%Src & HR & Variability & X-ray & $\rm M_{V}$ & $V-I$ & Range of & $\log(f_{x}/f_{o})$ & Multi- \\
%No. & Class & & Spectrum & & & $ $V-I$ & & wavelength \\
%\hline

T1 & XRB  & LONG & DISKBB & $-2.64$ & 0.72(a) & $[0.18, 0.72]$  & $[1.77, 2.06]$ (AGN)    & $-$       \\
   &      &      &        & $-2.33$ & 0.18(b) &  		&		       &	 \\
   &      &      &        & $-1.93$ & 0.62(c) &  		&		       &	 \\
T2 & XRB  & LONG & PL     & $-1.22$ & 0.55(a) & $[-0.82, 1.05]$ & $[2.79, 3.20]$ (AGN)    & FUV? (BWF15)    \\
   &      &      &        & $-1.01$ & $-0.45$(b)&  		&		       &	 \\
   &      &      &        & $-0.92$ & 0.46(c) &  		&		       &	 \\
T3 & SOFT & $-$    & $-$  & $-3.22$ & $-0.23$(a)& $[-0.23, 1.08]$ & $[0.18, 0.71]$ (HMXB)   & $-$       \\
   &      &      &        & $-2.35$ & 1.08(b) &  		&		       &	 \\
   &      &      &        & $-2.12$ & 0.55(c) &  		&		       &	 \\
T4 & SOFT & $-$    & $-$  & $-5.10$ & 0.99(a) & $[-0.40, 1.46]$ & $[-0.83, 0.55]$ (HMXB)  & $-$       \\
   &      &      &        & $-3.86$ & 0.48(b) &  		&		       &	 \\
   &      &      &        & $-3.68$ & $-0.38$(c)&  		&		       &	 \\
T5 & ABS  & LONG & $-$      & $-$ & $-$ & $-$           & $-$                    & $-$       \\
T6 & ABS  & LONG & $-$      & $-3.45$ & -0.15(a)& $[-0.15, 1.04]$ & $[0.83, 1.49]$ (HMXB?)  & $-$       \\
   &      &      &        & $-2.93$ & 0.79(b) &  		&		       &	 \\
   &      &      &        & $-2.25$ & 0.27(c) &  		&		       &	 \\
T7 & SOFT & $-$    & BBODY  & $-$ & $-$ & $-$           & $-$                    & $-$       \\
T8 & XRB  & $-$    & $-$      & $-2.84$ & 0.79(a) & $[-0.10, 0.79]$ & $[0.86, 1.36]$ (HMXB?)  & $-$       \\
   &      &      &        & $-2.40$ & 0.09(b) &  		&		       &	 \\
   &      &      &        & $-2.37$ & $-0.10$(c)&  		&		       &	 \\
T9 & SOFT & $-$    & $-$      & $-$ & $-$ & $-$           & $-$                    & $-$       \\
T10 & SNR & $-$    & PL     & $-3.61$ & 0.77(a) & $[-0.19, 1.41]$ & $[-0.27, 0.58]$ (HMXB)  & $-$       \\
   &      &      &        & $-3.36$ & 0.79(b) &  		&		       &	 \\
   &      &      &        & $-3.30$ & 1.16(c) &  		&		       &	 \\
T11 & ABS & LONG & PL     & $-$ & $-$ & $-$           & $-$                    & FUV? (BWF15)    \\
T12 & XRB & $-$    & PL     & $-$ & $-$ & $-$           & $-$                    & $-$       \\
T13 & ABS & $-$    & PL     & $-$ & $-$ & $-$           & $-$                    & $-$       \\
T14 & SOFT& $-$    & PL     & $-$ & $-$ & $-$           & $-$                    & $-$       \\
T15 & XRB & $-$    & $-$    & $-$ & $-$ & $-$           & $-$                    & Optical (SRW06) \\

%\enddata
\hline
\end{tabular} 
\tablecomments {(1) Source number; (2) Source classification based on the HRs; (3) Observed X-ray variability; (4) Best-fit X-ray spectral model when $\chi^2$ statistic is used; (5)--(6) The absolute $V$ magnitudes and $V-I$ colors of the brightest objects inside the \xmm{} error circle and their labels are given in bracket; (7) Range of $V-I$ color of the optical sources in the error circle; (8) Range of logarithmic X-ray--to--optical flux ratio of the optical sources and the likely class given in bracket; (9) Multi-wavelength information available from the literature (See the text for more details).}
\label{class}
%\end{deluxetable}
\end{table*}

\section{Discussion}

\subsection{BH and NS Candidates}

In addition to T1, which has been identified as a candidate BH
XRB (JW15), T2 is another bright source and was detected in the XMM1, XMM2 
and 2001 \chandra{} observations. However it completely disappeared in the 
long 127 ks \xmm{} observation. The analysis showed that T2 changed its 
luminosity by at least two orders of magnitude from $\sim 10^{38}$ 
to $<10^{36}~\rm erg~s^{-1}$. The HRs derived from the \xmm{} and \chandra{} 
observations classified this source as an XRB system and it exhibited 
strong short-term variability in these observations. The spectrum of T2 is 
better described by an absorbed power law with $\Gamma \sim 1.7$, 
indicating that the source was likely in the hard state \citep{Rem06}. 
The hard state is observed in BH and NS LMXBs \citep{Van94} only at 
luminosities $\leq 10\%$ Eddington luminosity $L_{Edd}$ \citep{Gla07}. 
Given its X-ray luminosity of $\sim 1.7 \times 10^{38}~\rm erg~s^{-1}$ 
in 0.3--8 keV band, which is $90\%\rm L_{Edd}$ for a $\rm 1.4\ \Msun$ NS,
we suggest that T2 is likely a BH XRB exhibiting a hard state behavior
during an outburst in 2001. 

The canonical model for BH XRB sources, namely a combination of power law 
and disc blackbody components, marginally improved the spectral fit for T2. 
The resulting inner disc temperature is unusually low ($\sim 0.14$ keV) 
indicating the presence of a cooler disc, which accounts for $< 35 \%$ of 
the total observed X-ray flux. However, the $\Delta \chi^2 \sim 4$ 
improvement for T2, for two extra degrees of freedom was not significant. 
Hence the presence of the soft component is tentative. The spectral 
modeling with double thermal components also 
marginally improves the fit compared to the single component model. For such
a case, the analysis still suggests a BH system, since the disc blackbody 
temperature and 2--10 keV disc blackbody luminosity of T2 do not fall in 
the NS LMXB parameter space. 

%(\#8, \#10, \#44, \#51, \#6 and \#37)
%(T12, T13,   T6,   T8,   T11 and T5)
Six sources (T5, T6, T8, T11, T12 and T13) are likely to be in either 
a hard state or a thermally dominated state. The X-ray luminosities of 
all these sources are $\leq 10\%$ Eddington luminosity for either NS or BH
primaries. This low accretion is observed in NS and BH XRBs in the hard state 
and suggests that the accretors of these binaries may be NS or BHs. However, there is 
no significant difference in the spectral fits and hence we cannot rule out 
a thermally dominated state from these low quality data. These sources could 
be NS or BH XRBs, and deeper observations are required for further 
classification.

\subsection{Super Soft Sources}
SSSs are classically defined as X-ray sources with temperature $\sim 100$ eV 
\citep{Dis03} and typical X-ray luminosities between $\sim 10^{36}$ and 
$10^{38}~\rm erg~s^{-1}$. The steady nuclear burning of matter accreted onto 
the surface of a white dwarf \citep{van92} seems to be the most promising 
explanation for majority of these sources. SRW06 identified seven SSSs, 
which is one-sixth of total X-ray population in NGC 55. We identified five 
transient sources, which are soft sources on the basis of their HRs. 
Their soft emission is modeled by a simple absorbed blackbody 
with temperatures of $\sim 50 - 180$ eV. The cool emission indicates 
that these sources are most likely WD systems. The HRs of three 
sources are consistent with those of ``typical'' SSSs. 
Due to the given transient nature of these sources, they 
are likely to be classical novae which are identified as the major class of 
SSSs in the galaxies M31 and M33 \citep{Pie05}. The X-ray emission from 
these systems arises due to the episodic thermonuclear burning of matter on 
the surface of the WDs, which leads to the transient nature \citep{Pie07}.      

%4 - snr  
%\#4 -> T10

We found that only one source (T10) had HRs consistent with those of SNRs,
although the uncertainties on the HRs, particularly on HR2, are quite large.
If we consider the thermal plasma model as a ``standard SNR'' model, 
the best fit plasma temperature of $\sim 0.2$ keV is lower than that of
the SNR candidates in M33 and the unabsorbed 0.5--2 keV luminosity is 
higher than the luminosities observed for the SNRs in M33 \citep{Lon10}. 
The blackbody fit rather provided a meaningful result, with temperature 
($kT \sim 230$ eV) consistent with QSSs. The emission from QSSs is too hot 
to be from WDs and the alternative physical systems include NS or BH 
binaries \citep{Dis10} and SNRs \citep{Ori06}.  

%Background sources

\subsection{Possible Optical Counterparts}

We searched the optical fields of 7 transient sources that are available 
from {\it HST} observations, which provide constraints on the source types for them. 
Although each \xmm{} source position contains multiple optical sources, 
if one of the sources is the counterpart, the X-ray--to--optical flux 
ratios suggest most of the transients as HMXBs, which is supported by 
the classification based on all available data. 
Further {\it HST} observations of the fields can identify the true
counterparts by detecting variables among the optical sources.

We also cross-correlated the sources with publicly available 
multi-wavelength catalogs. Three of them, T2, T11 and T15, 
has a counterpart identified in the multi-wavelength data, 
but the positions for the first two are not consistent with 
that from our \xmm{} analysis. T15 could be either a NS or a BH XRB 
based on the spectral properties, but it is optically identified as a
background AGN from multi-wavelength catalogs. SRW06 estimated 10--15 
background objects in the $\rm D_{25}$ region of the galaxy. We note that
some of the new sources identified in the long 127 ks observation have 
a color consistent with the absorbed category and they may be background 
objects. We estimated the number of background AGNs in the $\rm D_{25}$ region of 
NGC 55 using the log$N$--log$S$ relation reported in \cite{Cam01}, and 
the chance probability for detecting a background AGN inside the $\rm D_{25}$ 
region is $\sim 26\%$.

%{\bf According to the log$N$--log$S$ distribution in \cite{Cam01},
%the chance probability for detecting a background AGN inside the $\rm D_{25}$ 
%region is $\sim 26\%$.}

\subsection{Comparison to Other Galaxies}
From X-ray perspective, NGC 55 is a typical Magellanic-type dwarf galaxy, 
which has a similar structure to that of LMC \citep{Dev61}. The X-ray 
properties of NGC 55 (SRW06), such as the total mass, neutral hydrogen mass, 
star formation rate (SFR), diffuse emission component, and discrete point 
source luminosities, illustrate the striking similarity with other nearby 
Magellanic systems, NGC 4449 and LMC \citep{Wan91, Vog97, Sum03}. 
The bright X-ray sources in these galaxies are spatially associated with 
star forming regions and broadly follow the correlation between the X-ray 
luminosity and SFR established for active systems \citep{Ran03}. 

The transient source population has been extensively studied in 
our Milky Way \citep[][and many others]{Jai01, Tom05, Mcc06}, but 
relatively little work has been conducted on the extragalactic siblings. 
Due to the proximity, most extragalactic studies of transient sources 
concentrated on the Magellanic Clouds \citep[MCs; ][]{Kah96, Coe01} and 
M31 \citep[][and references therein]{Wil06}. A known population of 
transient sources in the MCs include XRBs and SSSs \citep{Kah08, Hab12, Stu13}. 
The sources identified as LMXB transients are rare in Magellanic Clouds, 
but the Small Magellanic Cloud hosts a large 
number of high-mass X-ray binaries with a possible B/Be type companion \citep{Kah96, Hab04}. 
Approximately 25--35 SSSs (including candidates) are observed in MCs, 
which are associated with magnetic cataclysmic variables, close 
binary SSSs (likely white dwarf and Be X-ray binaries), 
and classical post-nova \citep{Kah06, Kah06a, Kah08, Stu13}, implying 
that the SSS population is an essential component in the Magellanic systems. 
Such effort has not been taken for NGC 4449. 
In Table \ref{comp}, we summarize the comparison of transient properties 
of NGC 55 with MCs. Since MCs are frequent targets of many different X-ray 
observatories, a large number of transient sources were identified with 
luminosity ranges of $\sim 10^{34} - 10^{39} \rm erg~s^{-1}$ compared to 
NGC 55. However, the fractions of transient types in NGC 55 differ by 
no more than a factor of three with MCs. The marginal difference in 
the fractions of transient types may be due to observational effects such 
as effective extinction and the luminosity threshold of the available 
observations. This difference could be better investigated through 
continuous monitoring observations of NGC 55 as it is done for the MCs.
The similarity in the transient and SSS populations, along with the X-ray 
properties of NGC 55 with other nearby Magellanic systems mentioned above, 
supports NGC 55 as a typical Magellanic system.

%This marginal difference could be explained by the better sampling of NGC 55 like MCs.

\begin{table}
\tabletypesize{\small}
\tablecolumns{3}
\setlength{\tabcolsep}{4.0pt}
\tablewidth{240pt}
	\caption{Comparison of transient properties of NGC 55 with MCs}
 	\begin{tabular}{@{}lcc@{}}
	\hline
\colhead{Property} & \colhead{NGC 55} & \colhead{$\rm MCs^a$} \\
\hline

%\begin{deluxetable}{lcr}
%\tabletypesize{\small}
%\setlength{\tabcolsep}{6.0pt}
%\tablewidth{280pt}
%\tablecaption{Comparison of transient properties of NGC 55 with MCs}
%\startdata
%\hline
%\hline
%Property & NGC 55 & $\rm MCs^a$ \\
%\hline

Number of Transients  & $15$           & $64$ \\
Fraction of HMXBs     & $27\%$           & $58\%$ \\
Fraction of SSSs      & $33\%$           & $25\%$ \\
$\rm L_{X}$ ($\rm erg~s^{-1}$)	      & $0.15 - 16.6 \times10^{37}$           & $5\times10^{34} - 3\times10^{39}$ \\
$\rm L_{X}$ of SSSs ($\rm erg~s^{-1}$)	      & $1.5 - 2.7 \times10^{36}$           & $6.7\times10^{34} - 3\times10^{39}$ \\
%\enddata 
\hline
\end{tabular} 
\tablecomments {$^a$ \citet{Kah96, Hab99, Hab99a, Kah99, Kah08, Coe12, Li12} and references there in.}
\label{comp}
%\end{deluxetable}
\end{table}

\section{Summary}
We have presented a study of the transient X-ray source population in the 
Magellanic-type galaxy NGC 55. We analyzed the archival \xmm{} and \chandra{} 
observations of NGC 55 and identified 15 candidate transient sources in 
the $\rm D_{25}$ region of the galaxy. Their X-ray luminosities are in
the range of $10^{36}$ to $\sim 10^{38}$\,erg\,s$^{-1}$. The high
sensitivities of the archival observations indicate a flux change 
of 1--2 orders of magnitude for some of these sources. The X-ray colors of 
these sources suggest that the transient population is dominated by XRBs. 
We performed detailed spectral and timing analysis of the data for
two bright transient X-ray sources in the $\rm D_{25}$ region of the galaxy. 
The strong short-term variability seen in the sources, and the X-ray colors 
support the speculation of their high X-ray luminosity arises from an XRB 
system with a BH as the primary accretor. Six sources have spectral 
properties consistent with NS and BH XRBs accreting at both normal and 
high accretion rate. The transient population also seems to contain a sizable 
SSS component, perhaps one-third of the detected transient sources. 
The soft emission from them indicates that they are like white dwarf binaries. 
Optical analysis of archival {\it HST} data provides constraints 
on the likely nature for the sources and most of them are possible HMXBs. The cross-correlation with 
multi-wavelength catalogs identified one source as a possible background object, which is 
classified optically as a galaxy. From the spectral analysis a variety of 
properties have been determined, revealing the heterogeneous nature of 
transient sources in NGC 55. Our study suggests that the Magellanic-type 
galaxies could be a potential factory of transients, and comparative studies 
with other Magellanic-type galaxies and continuous monitoring at X-ray (multi)
wavelengths will help us understand the physical nature of various types
of high-energy objects that comprise the transient population.

\section*{Acknowledgements}
We thank the anonymous referee for the helpful comments and recommendations 
that improved this manuscript. This work has made use of data obtained from the 
High Energy Astrophysics Science Archive Research Center (HEASARC), 
provided by NASA's Goddard Space Flight Center. This research was 
funded by Chinese Academy of Sciences President’s International 
Fellowship Initiative (CAS PIFI, Grant No. 2015PM059), and was supported in 
part by the Strategic Priority Research Program 
``The Emergence of Cosmological Structures" of the Chinese Academy 
of Sciences (Grant No. XDB09000000).

%\bibliography{transient}

\end{document}